\documentclass[fleqn,usenatbib]{mnras}

\usepackage{newtxtext,newtxmath}

\usepackage[T1]{fontenc}
\usepackage{ae,aecompl}

\usepackage[dvipsnames]{xcolor}

\usepackage{graphicx}	
\usepackage{amsmath}	
\usepackage{listings}
\usepackage{soul}

\newcommand{\fink}{{\sc Fink}}

\title[Satellite glints in ZTF]{Impact of satellite glints on the transient science on ZTF scale}

\author[Karpov, Peloton]{Sergey Karpov$^{1,3,4}$\thanks{E-mail:karpov@fzu.cz},
Julien Peloton$^{2}$\thanks{E-mail:julien.peloton@ijclab.in2p3.fr}
\\
$^{1}$CEICO, Institute of Physics, Czech Academy of Sciences, Prague, Czech Republic \\
$^{2}$Universit{\'e} Paris-Saclay, CNRS/IN2P3, IJCLab, Orsay, France\\
$^{3}$Special Astrophysical Observatory, Russian Academy of Sciences, Nizhniy Arkhyz, Russia \\
$^{4}$Kazan Federal University, Kazan, Russia \\
}

\pubyear{2022}

\begin{document}
\label{firstpage}
\pagerange{\pageref{firstpage}--\pageref{lastpage}}
\maketitle

\begin{abstract}
Thousands of active artificial objects are orbiting around Earth 
along with much more
non-operational ones resulted from human activity -- derelict satellites or rocket bodies, collision debris, or spacecraft payloads. Significant part of them are uncatalogued, and they represent potential dangers for space instruments and missions. They also impact observations of the sky by ground-based telescopes by producing a large number of streaks polluting the images, as well as generating false alerts hindering the search for new astrophysical transients. While the former threat for astronomy is widely discussed nowadays in regard of rapidly growing satellite mega-constellations, the latter one -- false transients -- still lacks attention on the similar level. 

In this work we assess the impact of satellite glints -- rapid flashes produced by reflections of a sunlight from flat surfaces of rotating satellites -- on current and future deep sky surveys such as the ones conducted by the Zwicky Transient Facility (ZTF) and the Vera Rubin Observatory Legacy Survey of Space and Time (LSST). 
For that, we propose a simple routine that detects, in a single exposure, a series of repeated flashes along the trajectories of otherwise invisible satellites, and describe its implementation in \fink\ alert broker. Application of the routine to ZTF alert stream revealed about 73,000 individual events polluting 3.6\% of all ZTF science images between November 2019 and December 2021 and linked to more than 300 different glinting satellites on all kinds of orbits, from low-Earth up to geostationary ones. The timescales of individual flashes are as short as $0.1$--$10^{-3}$ seconds, with instant brightness of 4--14 magnitudes, peak amplitudes of at least 2--4 magnitudes, and generally complex temporal patterns of flashing activity. We expect LSST to see much more such satellite glints of even larger amplitudes due to its better sensitivity.

\end{abstract}

\begin{keywords}
surveys -- methods: data analysis -- transients  -- space vehicles -- light pollution
\vspace{-1.5\baselineskip}  
\end{keywords}


\section*{Introduction}
\label{sec:introduction}

Modern large-scale time-domain sky surveys, like upcoming Vera C. Rubin Observatory Legacy Survey of Space and Time \citep{lsst} and ongoing Zwicky Transient Facility \citep{Bellm_2018} offer an unprecedented possibility of not only detailed study of already known classes of transient objects, like variable stars, novae, supernovae, tidal disruption and microlensing events, AGNs and distant Solar System objects, but also of discovery of potentially new and exciting classes of astrophysical transients, especially in the still poorly studied region of shortest time scales. Indeed, despite long history of various experiments aimed towards investigations of transients on time scale of fractions of seconds to a few seconds \citep{schaefer_1987, rotse_2002, karpov_2010, karpov_2019, richmond_2020, tingay_2020, arimatsu_2021}, they had only a limited success in detecting initial phases of gamma-ray bursts \citep{racusin_2008, karpov_grb_2017, zhang_2018}, in part due to quite limited depth of these surveys.

On the other hand, the zoo of rapid optical transients is potentially extremely diverse, spanning from short intense flares on red dwarf stars to gamma-ray bursts and fast radio bursts \citep{frb}.
And while modern and upcoming surveys like ZTF and LSST have insufficient temporal resolution and cadence to properly sample the light curves of such events, due to extreme sensitivity they still may detect both their peaks and afterglows, as it was demonstrated by the recent ZTF discovery of orphan GRBs \citep{ztf_orphans, ztfrest}.

The search for rapid optical transients on a sub-second time scales is significantly complicated by the large background of artificial events -- satellite glints, or flares, caused by the sunlight reflection from solar panels or rotating antennae onboard an artificial satellites orbiting the Earth. Such events have both sub-second duration and are bright enough to be apparent even in smaller-scale wide-field surveys \citep{mmt_flashes, nir_2020, corbett_2020}. 
Even in deep surveys with lower temporal resolution, where satellites typically produce extended streaks and trails easily detectable by their morphology, such rapid flashes still generate point sources\footnote{These events are also much fainter when observed with longer exposures -- e.g the flash with 0.3 second duration and 10th mag peak amplitude will be seen as a 15th mag object in 30 second exposure image.} indistinguishable from stellar or extragalactic objects that may mimic e.g. gamma-ray bursts \citep{nir_2021}.

In this article we investigate the impact of such satellite glints 
on the transients detectable in the ZTF alert data stream using the data collected by the \fink\ broker\footnote{\url{https://fink-broker.org}} \citep{fink} between November 2019 and December 2021 (523 observational nights). We developed a simple methodology for searching for such events among millions of candidates, and successfully recovered tens of thousands of them. 

The approach we use is based on purely geometric analysis of the transients detected on individual exposure, and fills the gap between existing efforts of extracting slowly moving objects \citep{Masci_2019} and analyzing the trails caused by rapidly moving ones \citep{zstreak,deepstreaks,2022ApJ...924L..30M}.

The paper is organized as follows. In Section~\ref{sec:implementation} we describe the methodology for extracting the series of events caused by satellite glints -- tracklets -- from the ZTF data stream using \fink. Section~\ref{sec:properties} details the properties of detected tracklets and known satellites associated with them, as well as their temporal properties. Finally, in Section~\ref{sec:discussion} we assess the efficiency of detection of satellite glints with this method, compare their brightness with other optical observations of satellites, and discuss their possible the impact on future LSST survey. The reader will also find additional information in appendices, especially the instructions on how to access the tracklet data in \fink, in Appendix~\ref{sec:science-portal}.
 
\section{Detection of satellite glints with \fink}
\label{sec:implementation}

\subsection{Alert ingestion and data selection}
\label{sec:data-ingestion}

\begin{figure}
    \centering
    \centerline{
        \resizebox*{1\columnwidth}{!}{\includegraphics{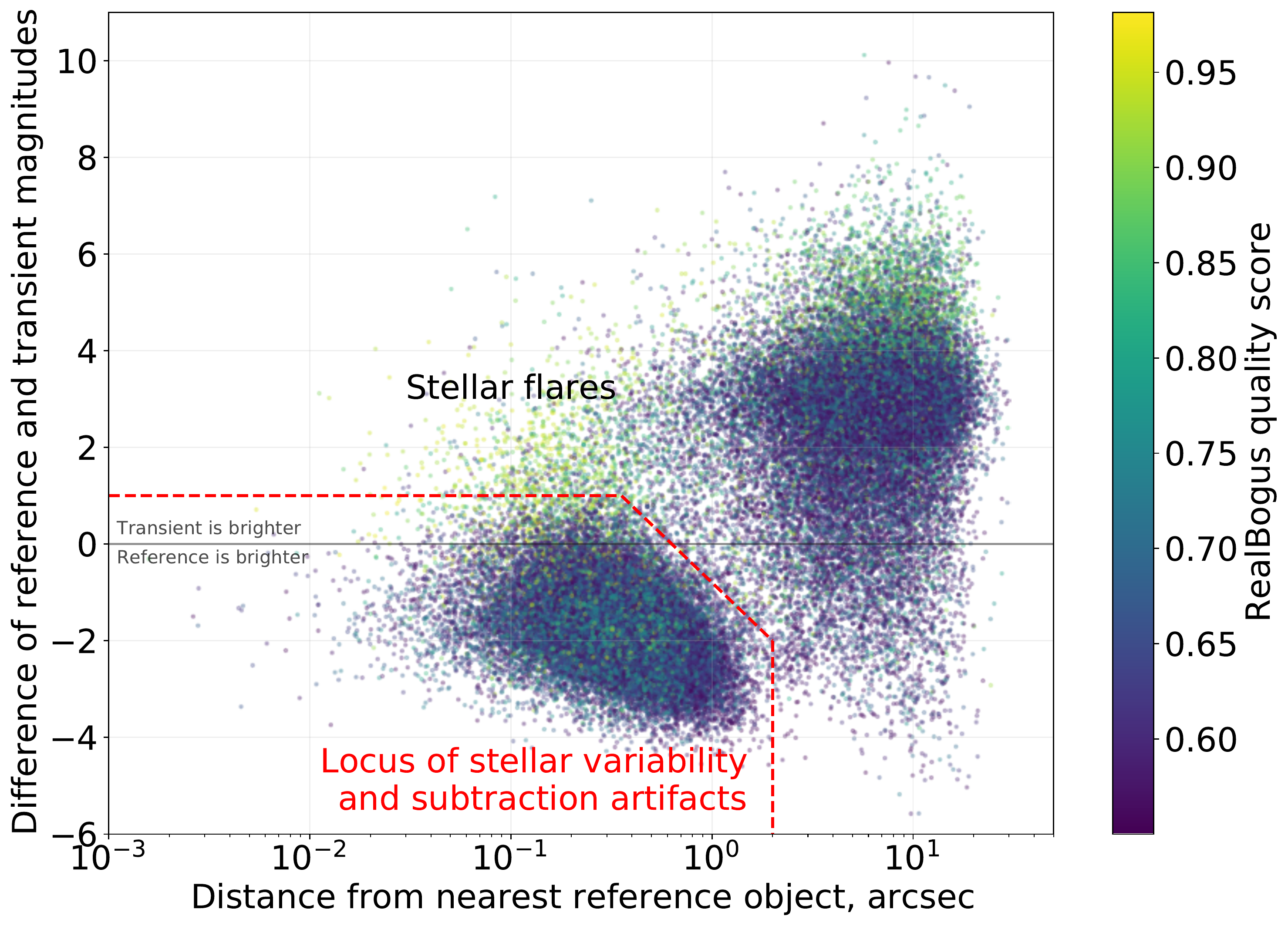}}
    }
    \caption{The difference between transient magnitude measured on difference image (\texttt{magpsf} field in ZTF AVRO scheme, corresponding to PSF-fitting measurements) and the magnitude of nearest object from reference catalogue (\texttt{magnr} field, also corresponding to PSF-fitting measurements) versus the distance to this nearest reference object for all events corresponding to non-repeating positive detections not coincident with Solar System objects. 
    The value of RealBogus quality score (\texttt{rb} field) is represented by the color of dots. Clearly visible cloud of lower-quality events in the lower part of the plot corresponds primarily to subtraction artefacts on stationary non-variable objects, as well as to occasional stellar variability like lower-amplitude stellar flares. Red dashed lines mark the extent of this ``stellar locus'' as used for rejection of these events in Section~\ref{sec:initial_selection}. Second cluster at (\texttt{magnr}-\texttt{magpsf})$~\sim3$ and \texttt{distnr}$~>2$ corresponds to random associations between faintest of noise events and faintest reference stars, as ZTF reference images are approximately 3 magnitudes deeper than individual exposures.}
    \label{fig:locus}
\end{figure}

\fink\ is a community driven broker project that processes time-domains alert streams and connects them with follow-up facilities and science teams \citep{fink}. The broker filters, aggregates, enriches, and redistributes incoming data streams in real-time. \fink\ will process the full stream of transient alerts from the Vera C. Rubin Observatory during its operations, together with other teams\footnote{\url{https://www.lsst.org/scientists/alert-brokers}}.

While preparing for the LSST alert data processing, \fink\ is being applied to the ZTF public stream \citep{Bellm_2018} since late 2019. The ZTF alert stream is an unfiltered, 5-sigma alert stream extracted from difference images. The alert stream primarily includes events from flux transients, variables, and Solar System objects. Detectable streaks, such as aircraft and satellite streaks, as well as cosmic rays, are removed by the ZTF processing pipeline prior to generating alerts \citep{Masci_2019}. {\color{gray}Hence we do not expect to receive bright and rapidly moving objects, including a large portion of space debris, that leave clear tracks on the images flagged by the \texttt{CreateTrackImage} software \citep{2014PASP..126..674L}.
}

Each night, the broker ingests in real-time the alert packets sent by the telescope. Each packet includes information on the alert candidate such as the timestamp, position, magnitude estimate, or data calibration, along with image cutouts around the position of the alert candidate as well as its history over a month. The collected data between November 2019 and December 2021 represents about 4.7TB of compressed data per month for ZTF (about 120 millions individual alerts, across 523 observation nights). Among these alerts, \fink\ processed about 75 millions alerts that satisfy the quality criteria. These criteria ensure that the data are of sufficient quality to extract scientific information. We currently apply two quality criteria: (1) the real-bogus score \citep{2019PASP..131c8002M} must be above 0.55, and (2) there should be no prior-tagged bad pixels in a 5 x 5 pixel stamp around the alert position.

Alerts that satisfy the quality criteria are then processed by the \fink\ pipelines to extract scientific information and reveal the nature of the objects. First, all alerts are crossmatched against the SIMBAD database \citep{SIMBAD:2000} to associate alerts to known catalogued objects. \fink\ has also several science modules targeting a wide range of science topics: supernovae, kilonovae, gamma-ray bursts, microlensing events, solar system objects, ... These science modules use the information contained on each alert (e.g. photometry, cutout images, position) to try to assess the nature of the transient. Among the 75 millions processed alerts, the \fink\ classification leads to: 35\% of alerts found in the SIMBAD database, 0.5\% of alerts flagged by at least one \fink\ science module (supernovae, kilonovae, microlensing events, gamma-ray bursts), 14\% alerts flagged as known Solar System objects found in the Minor Planet Center (MPC) database\footnote{\url{https://www.minorplanetcenter.net/}}, 0.4\% alerts flagged as new Solar System object candidates, and 50\% alerts remain without clear classification at this stage.

\subsection{Selection of candidate events}
\label{sec:initial_selection}

\begin{lstlisting}[language=Python, frame=single, float=t, caption={The final set of cuts applied to the alert stream before geometric extraction of tracklets. The parameters here directly correspond to the names of fields in ZTF AVRO alert schema, and the conditions are represented with pseudo-code.}, label=listing:cuts]
# Reject alerts with multiple appearances within 1.5 arcseconds
ndethist == 1 
# Reject alerts with MPC objects within 30 arcseconds
ssnamenr == 'null'
# Consider only positive detections on difference frame
isdiffpos == 't' 

# Reject locus of stellar variability and subtraction artefacts
(magnr - magpsf > 1.0)
OR (magnr - magpsf > 4*(log10(distnr) + 0.2))
OR (distnr > 2)
\end{lstlisting}

In order to target the events linked with artificial satellites, we started with a set of all non-repeating alerts not connected to known stationary objects, as well as to known moving Solar System objects. To reject the latters, we excluded every alert that is within 30 arcseconds of current positions of objects from Minor Planet Center (MPC) database. 

To exclude the events related to stationary objects we located and excluded the ``stellar locus'' on the plane defined by the distance of the transient to the closest object in the reference catalogue and the difference of their magnitudes as shown in Figure~\ref{fig:locus}. This locus contains the alerts related to both occasional stellar variability (i.e. small amplitude stellar flares) and -- most important for our analysis -- the image subtraction artefacts. The characteristic features of the latters are their innate proximity, within couple of typical PSF sizes, to the objects visible on reference frame (and thus populating the reference catalogue), and the brightness not exceeding the one of reference object. This locus contains on average 50\% percents of isolated non-repeating alerts not connected to known Solar System objects.

\begin{figure}
    \includegraphics[width=\linewidth]{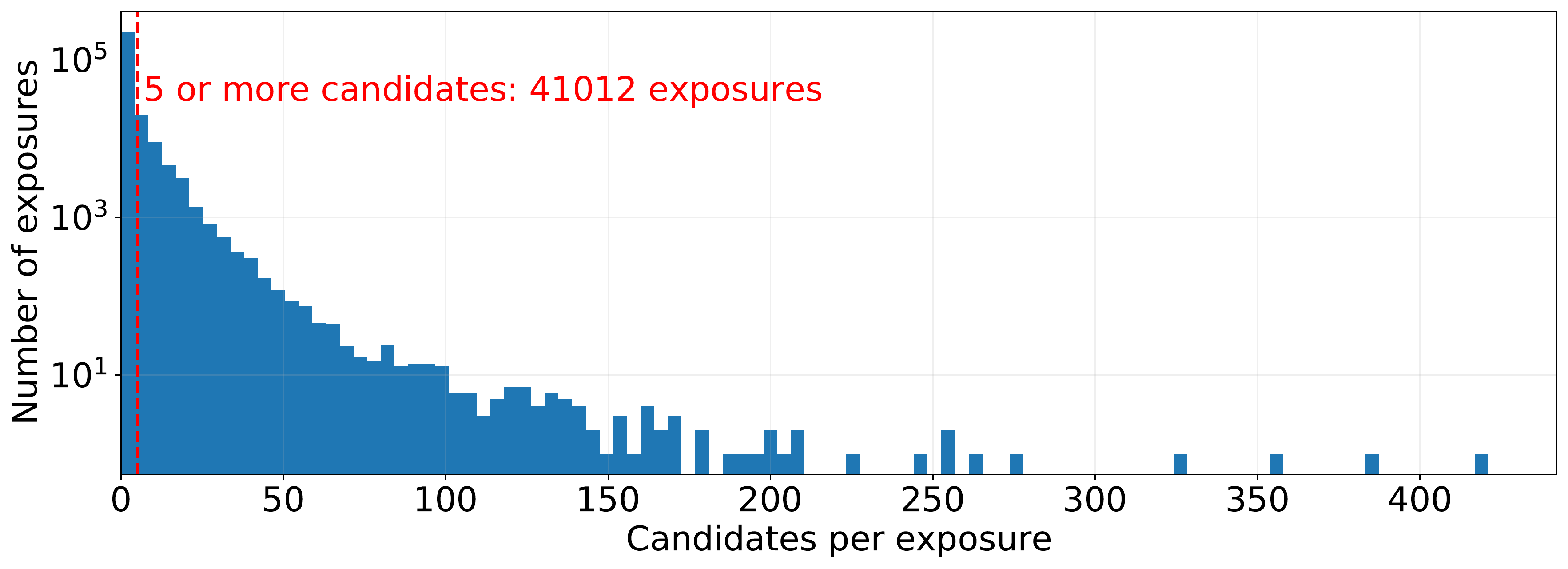}
    \caption{Histogram of number of candidate alerts passing all the cuts as defined in Section~\ref{sec:initial_selection}, per exposure. There is a total of 638,350 candidates among 113,422 exposures (41,012 exposures with 5 or more candidates).}
    \label{fig:nperframe}
\end{figure}

\begin{figure}
    \centering
    \centerline{
        \resizebox*{1\columnwidth}{!}{\includegraphics{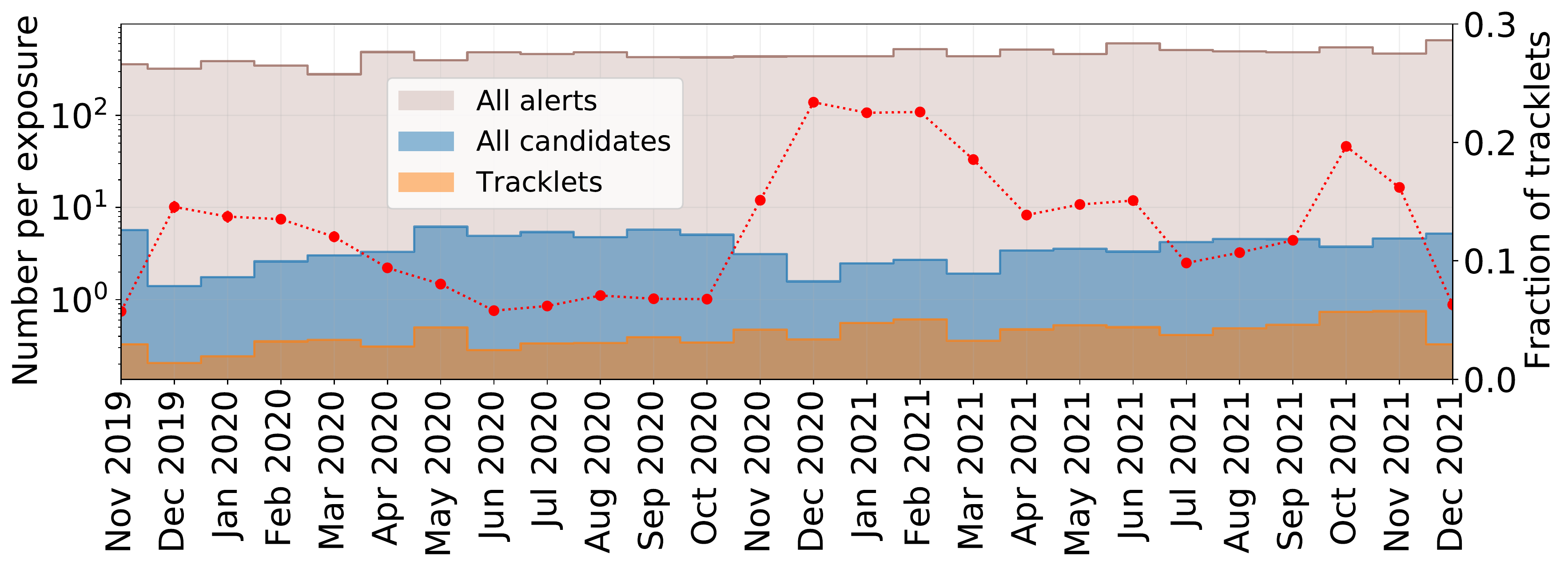}}
    }   
    \caption{
    Mean rate of all alerts processed by \fink\ per ZTF exposure along with the same for all candidate events passing quality cuts as defined in Section~\ref{sec:initial_selection}, as well as the ones belonging to tracklets detected by the algorithm described in Section~\ref{sec:tracklet_detection}. Also, the fraction of tracklet events among all candidates is shown with red circles. The data is binned on a per-month basis.
    }
    \label{fig:rate}
\end{figure}

The final set of cuts applied to the alert stream is defined solely using the parameters available in the alert packets and is shown in Listing~\ref{listing:cuts}. The statistics of events passing these criteria is shown in Table~\ref{tab:tracklet_selection}, and the histogram of amount of final candidates passing all criteria per ZTF exposure is shown in Figure~\ref{fig:nperframe}. From the 164,163 exposures acquired from November 2019 till December 2021, in total 113,422 (69\%) exposures contain
candidate events, and about 41,000 (25\%) of them -- 5 or more such candidates.
The rate of candidate events over time is shown in Figure~\ref{fig:rate}, and is sufficiently stable apart from some minor annual variations that are most probably due to variations of observing conditions.

\begin{figure}
    \centering
    \centerline{
        \resizebox*{1\columnwidth}{!}{\includegraphics{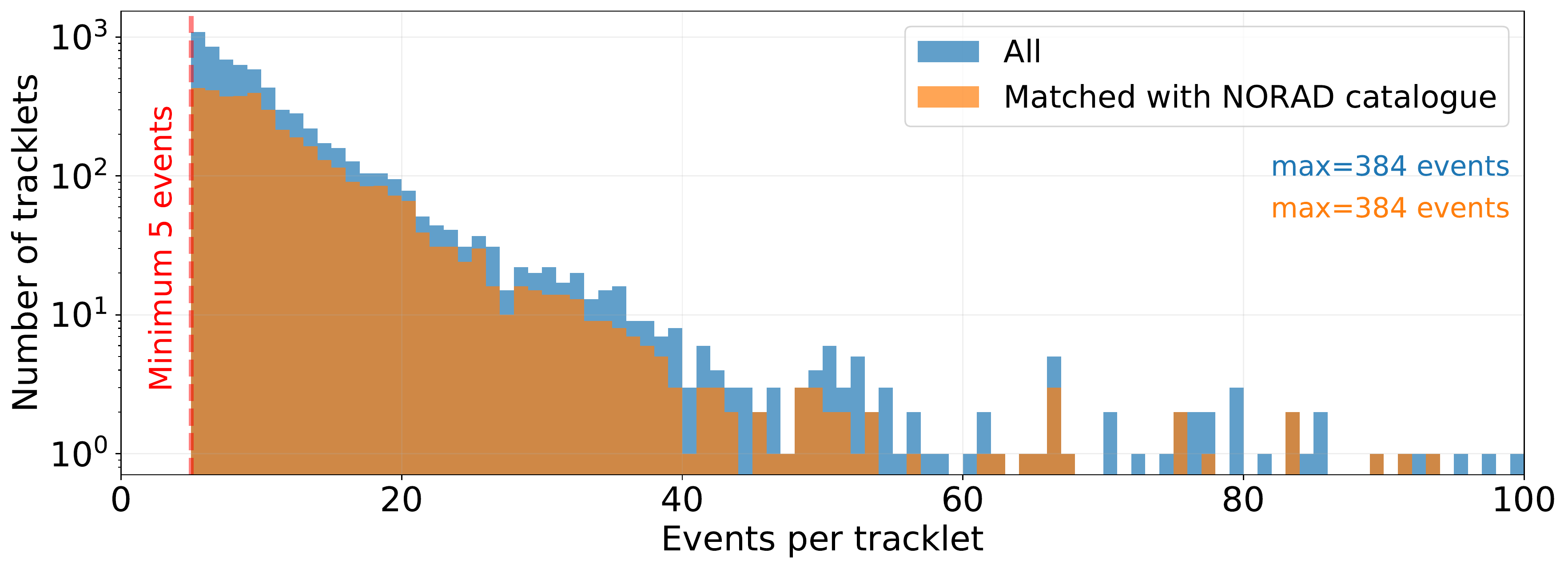}}
    }
    \centerline{
        \resizebox*{1\columnwidth}{!}{\includegraphics{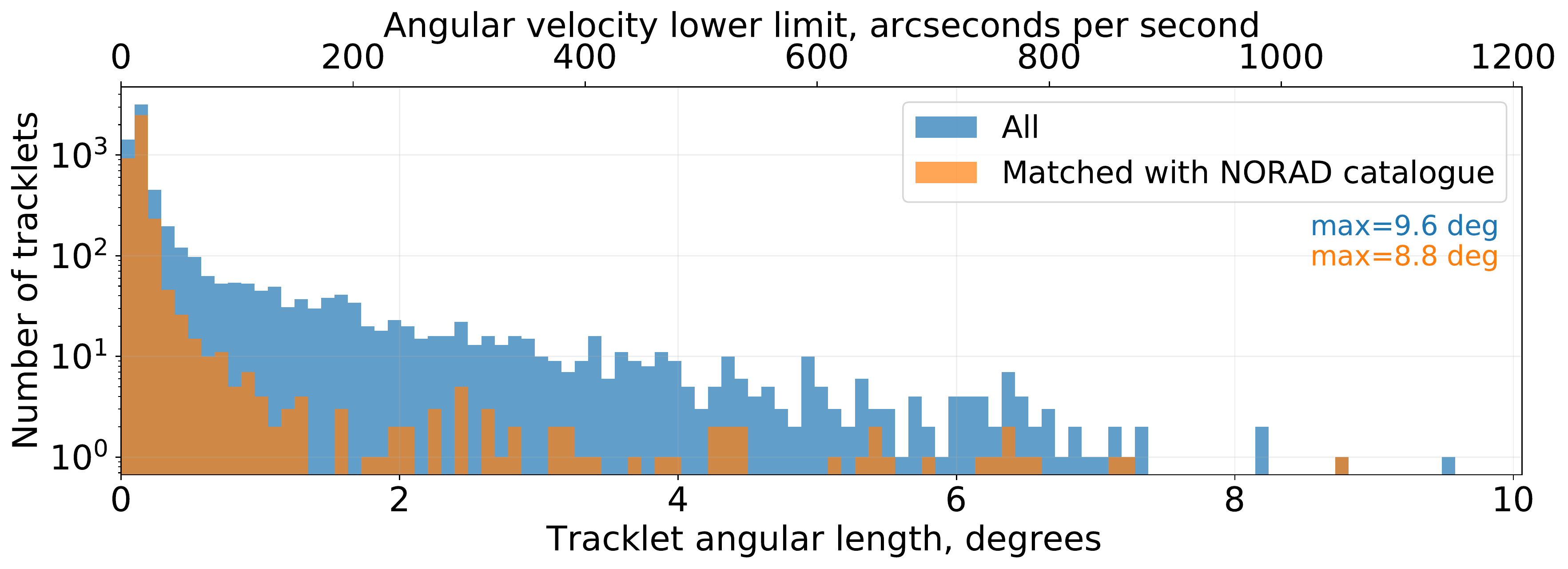}}    
    }
    \caption{Upper panel -- histogram of number of events per detected tracklet (blue histogram), limited to 100 events per tracklet. Lower panel -- angular length of detected tracklets, and corresponding lower limits on apparent angular velocities, assuming an exposure duration of 30 seconds (blue histogram). On both panels, the fraction of tracklets matched with passes of satellites from the public part of NORAD catalogue (see Section~\ref{sec:satellites_match}) is also shown in orange.}
    \label{fig:trackletlength}
\end{figure}

\begin{table}
    \begin{center}
        \begin{tabular}{lr}
        \hline\hline
        Alerts processed by \fink\   &  75,913,319 \\
        + single positive detection & 11,384,517  \\
        + no Solar System objects & 1,297,307 \\
        + outside stellar locus & 638,350  \\
        + belong to tracklets & 73,368  \\
        \hline
        Number of unique tracklets   & 6,450 \\
        \end{tabular}
        \caption{Statistics of events passing various criteria used to define the candidates for tracklet search on the period from November 2019 to December 2021. We show the number of alerts that pass each step, having applied that criteria over the remaining alerts from the previous stage. The last line shows the total number of detected tracklets over the entire period, which corresponds to about 12 detected tracklets or 140 alerts per observational night. The criteria are further described in Sections \ref{sec:initial_selection} \& \ref{sec:tracklet_detection}}
        \label{tab:tracklet_selection}
    \end{center}
\end{table}

\subsection{Detection of tracklets}
\label{sec:tracklet_detection}

As the apparent motion of artificial satellites, especially the ones on geostationary orbits, on the sky cannot be adequately approximated by a great circle over longer time intervals, and, on the other hand, the consecutive ZTF exposures are often acquired in quite distant sky directions, we decided to limit our analysis to detecting the sequences of events corresponding to the same satellite -- tracklets -- inside individual exposures only. Moreover, even for a typical ZTF survey exposure of 30 seconds only, due to its large simultaneous field of view (which is slightly less than 7x7 degrees), the motion of satellites can significantly deviate from the great circle on the sky.

Thus, we define the tracklet as several events detected on the same exposure and located along the same smooth curve on the sky.
We also require it to  contain at least 5 points in order to minimize false associations, corresponding to $10^{-11}$ probability of a random coincidence for 5 candidate events per exposure, and $10^{-5}$ -- for 50 candidates per exposure\footnote{After filtering, we have 638,350 candidates spread over 113,422 exposures, with a median at 3 candidates/exposure.}. As we do not have any temporal information associated with individual points of the tracklet, and therefore cannot exploit typical multi-frame object linking methods, in order to detect such features we developed a simple algorithm based purely on geometric positions of individual points as described below.

\begin{figure}
    \includegraphics[width=\linewidth]{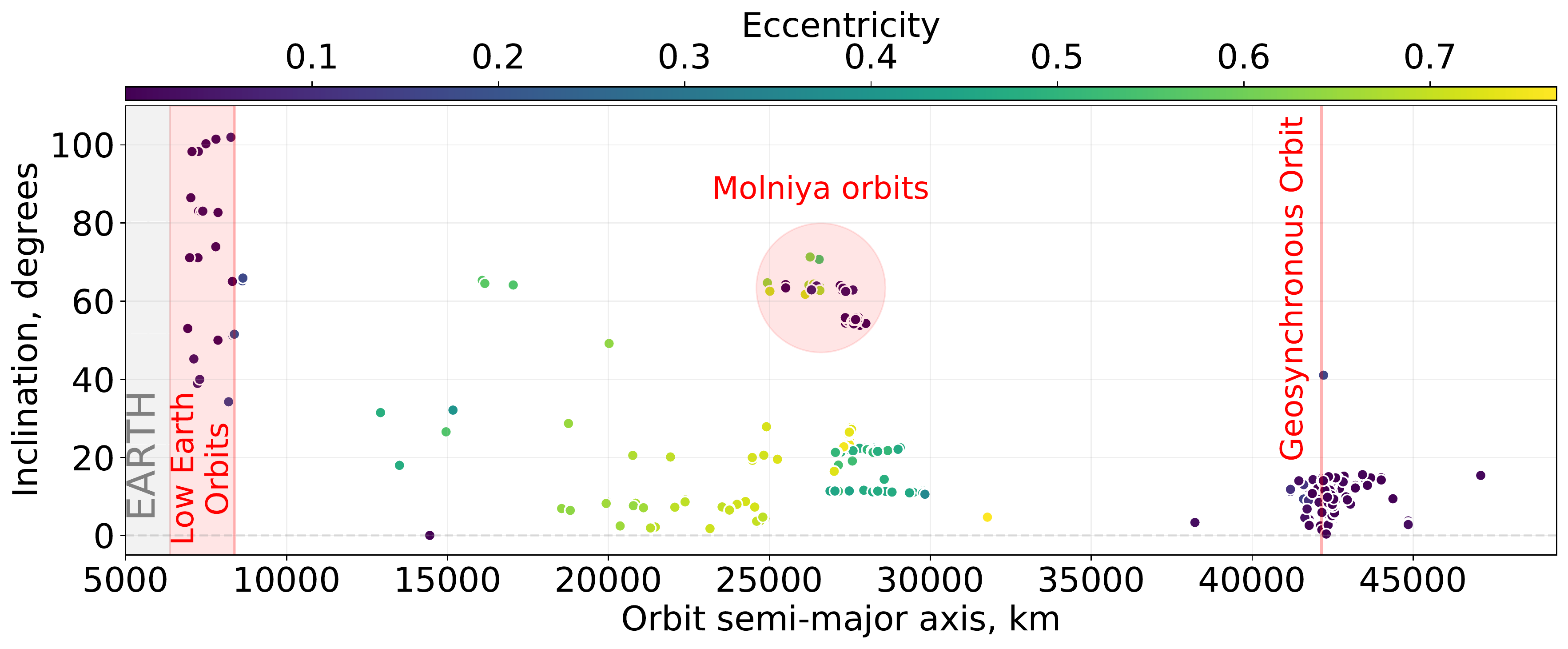}
    \caption{Orbital parameters for the satellites from NORAD catalogue matched with tracklets: inclination as a function of the orbit semi-major axis, color-coded by the eccentricity. We display three zones of interest: Low-Earth orbits (with semi-major axis below an altitude of 2000 kilometers), Molniya orbits (centered on a semi-major axis corresponding to half a sidereal day, 26,000 kilometers, and an inclination of 63.4 degrees), and geosynchronous orbits.
    Every point represent single association, so the same satellite may appear multiple times with different orbital elements, if it is observed multiple times, and its orbit gradually evolves.
    }
    \label{fig:orbits}
\end{figure}

First, for every pair of candidates from a given exposure separated by at least 10$''$, we constructed a great circle passing through both of them. Next, we computed the distances of every candidate from this great circle, and kept only the circles with at least 5 candidates closer than 5$''$ (which is 5 pixels on ZTF images, and should be enough to capture small systematic deviations from the great circle) to them, and their corresponding sets of points. Next, we iteratively fit the separation of the points from a great circle with a second order polynomial of a projected distance along the same circle, capturing new points not initially included into the set but accepting only points closer than 1$''$ (1 pixel) to the fit, until the fit converges or less than 5 points remains in the set (then it is discarded). Moreover, we specifically exclude from the set the first or last point, when ordered along the great circle, if their separation from the rest is larger than 10 times length of the rest of the set, thus requiring the set to be more or less continuous but allowing for some gaps in the track (e.g. due to non-detection of corresponding events).
Finally, we merge together the sets of points that have 3 or more points in common, allowing for some deviations from the great circle that can't be adequately represented by second order polynomial, but are still quite smooth -- in contrast to e.g. intersection of two different trajectories that will have no more than single common point.

The algorithm outlined above has been successfully applied to all historical alerts archived with \fink\ from 164,163 unique ZTF exposures covering the November 2019 -- December 2021 time interval, and as a result of its operation we detected 6,450 tracklets containing 73,368 candidates (11.5\% of all events passing selection cuts defined in Section~\ref{sec:initial_selection}). The tracklets are detected on 5,946 (3.6\%) unique exposures, and their properties are discussed in the next section.

\section{Properties of detected events}
\label{sec:properties}

As a result of operation of tracklet detection algorithm described in Section~\ref{sec:tracklet_detection}, we successfully detected large number of tracklets -- series of glints on a single exposure produced supposedly by the same satellite. Median number of events per tracklet is 9, with longest track containing 384 events, and their median angular length on the sky is 416$''$ (see Figure~\ref{fig:trackletlength}).

\subsection{Cross-identification with NORAD satellite catalogue}
\label{sec:satellites_match}

In order to characterize the physical properties of the objects behind these tracklets, we cross-matched them with the positions of known satellites from public NORAD catalogue published by the Joint Space Operation Center (JSpOC). As the orbital elements rapidly evolve in time, we used a sets of two-line elements (TLE) downloaded daily from  \url{https://www.space-track.org} for the analysis of tracklets detected on the same day. Then, we propagated these orbital elements to the times of individual exposures using {\sc Skyfield} package \citep{skyfield}. In order to associate the satellites with tracklets, we used the following criteria.

First, we excluded all satellites with the distance of their position at the start of exposure from the first point of a tracklet exceeding 20 degrees.
For the rest, we computed the normals to great circles formed by first and last points of a tracklet, and by the positions of a satellite at the start and end of the exposure. Then, we excluded all associations with the angle between these normals exceeding 1 degree, thus keeping only satellites with paths roughly co-linear on the sky to the tracklet. Finally, we computed the average distance of tracklet points to the great circle of a satellite path, and rejected the association if this distance is exceeding 300 arcseconds. Then we selected for every tracklet the satellite with smallest average distance, if any, and considered it a match.

\begin{figure}
    \centering
    \centerline{
        \resizebox*{\linewidth}{!}{\includegraphics{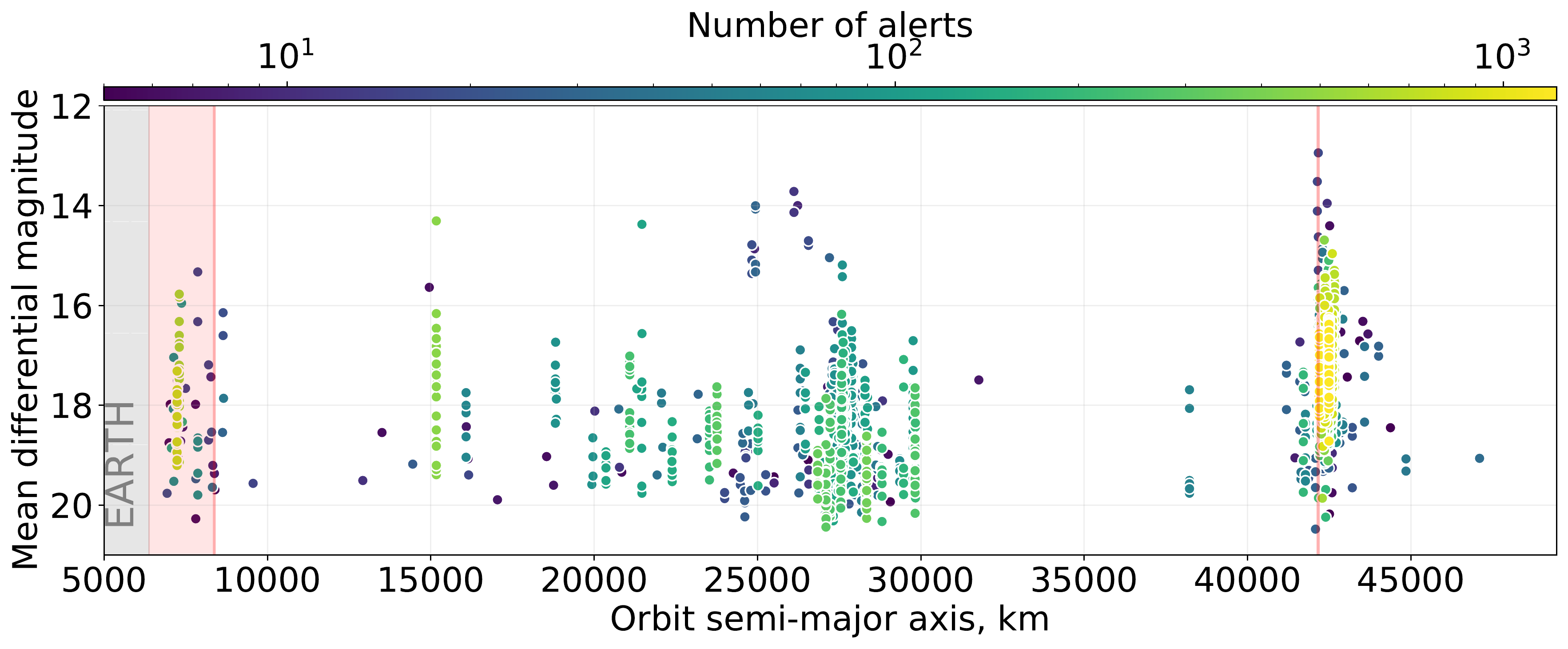}}
    }   
    \caption{Mean differential PSF magnitude of tracklet alerts as a function of associated satellite orbit semi-major axis. The marker color corresponds to the total number of alerts produced by a satellite. We also display two zones of interest: Low-Earth orbits and geosynchronous orbits,
    analogous to the ones shown in Figure~\ref{fig:orbits}.
    }
    \label{fig:nalerts}
\end{figure}

This way, we acquired matching satellites for 3,841 (60\%) of all tracklets, containing 45,387 (62\%) events. The fraction of matched tracklets does not depend on the number of points in the tracklet, but drops for the tracklets with longer arcs on the sky (see Figure~\ref{fig:trackletlength}) -- those correspond to lower-orbit objects with faster evolving orbits, making it more difficult to find an associated satellite for a tracklet.
In total, we associated the tracklets with 308 unique NORAD object IDs, with 94 of them observed for 10 and more (up to 96) times. For reference we list the objects that produce the most of tracklets in Table \ref{tab:sat_name}.

We combined their instantaneous orbital parameters derived from TLE with physical parameters and satellite status information from CelesTrak Satellite Catalogue\footnote{CelesTrak Satellite Catalogue is available online at \url{https://celestrak.com/satcat/}. It contains, in addttion to rough orbital parameters, also the information on satellite operational status, as well as its Radar Cross-Section value that may be used as a rough estimate of satellite size. The latter information is not available for the objects launched since 2017.}. Figure~\ref{fig:orbits} shows the distribution of orbital elements of these satellites. Several distinct populations are visible there -- from smaller objects on low-Earth orbits with significant inclinations and low eccentricity to larger ones on a geosynchronous orbits with low inclinations, as well as populations of Molniya-type ones and a set of semi-synchronous objects with small inclinations but large eccentricity. Among all 308 associated satellites, only 9 are marked as still operational (active) in CelesTrak Satellite Catalogue, while 126 are non-operational, and 170 have unknown status (but with 105 of them having 'DEB' suffix in the name suggesting it is a debris, 11 -- 'AKM', meaning that it is an apogee kick motor, and 11 more -- 'R/B', suggesting that it is a rocket body). Thus, we may conclude that the majority of glints (about 97\%) we detected is produced by inactive satellites.

Figure~\ref{fig:nalerts} also shows the mean brightness of the glints (represented by their differential PSF magnitudes) as a function of associated satellite semi-major axis, along with the number of alerts produced by an individual satellite. There is no obvious systematic trend in brightness (as it is most probably significantly affected by selection effects on the brighter end, and is strongly constrained by ZTF detection limit -- on the fainter one), 
however, the objects most frequently producing the glints are all situated on geostationary orbits, in agreement with Table~\ref{tab:sat_name}.

\begin{table*}
\begin{tabular}{rlccccccccc}
\hline
ID & Satellite name & Ntracklets & Nalerts & Status & $a$, km & RCS, m$^2$ & Arc, deg. & $P_{\mbox{m}}$, s. & $P_{\mbox{p}}$, s. & $\tau$, s. \\ 
\hline\hline

21964 & PALAPA B4 & 96 & 810 & Unknown & 42261 & 12.6 & 0.13 & 0.5 & 3 & 0.07 \\
24769 & BSAT-1A & 95 & 1223 & Inactive & 42492 & 15.8 & 0.12 & 0.4 & 1 & 0.07 \\
23314 & THAICOM 2 & 84 & 937 & Inactive & 42359 & 1.3 & 0.12 & 0.5 & 3 & 0.07 \\
14134 & PALAPA B1 & 77 & 1151 & Inactive & 42199 & 1.3 & 0.12 & 0.4 & 2 & 0.07 \\
23016 & GALAXY 1R & 74 & 727 & Inactive & 42463 & 8.3 & 0.12 & 0.7 & 3 & 0.07 \\
25312 & BSAT-1B & 72 & 794 & Inactive & 42492 & 12.6 & 0.12 & 0.4 & 2 & 0.07 \\
20402 & JCSAT 2 & 71 & 677 & Inactive & 42659 & 8.2 & 0.12 & 0.3 & 2 & 0.07 \\
14234 & ARABSAT 1DR (TELSTAR 3A) & 69 & 866 & Inactive & 42377 & 2.5 & 0.13 & 0.6 & 1 & 0.07 \\
22931 & THAICOM 1 & 68 & 671 & Inactive & 42474 & 1.0 & 0.12 & 0.5 & 3 & 0.07 \\
20193 & SIRIUS W (MARCOPOLO 1) & 67 & 891 & Unknown & 42473 & 2.0 & 0.12 & 0.4 & 1 & 0.07 \\

\multicolumn{11}{c}{--- // ---} \\

28556 & ARIANE 1 DEB & 1 & 5 & Unknown & 27013 & 0.1 & 0.051 & 0.7 & 1 & 0.2 \\
22911 & SOLIDARIDAD 1 & 1 & 5 & Inactive & 42164 & 12.5 & 0.12 &  & 4 & 0.07 \\
15386 & MARECS B2 & 1 & 5 & Inactive & 43429 & 3.2 & 0.11 &  & 3 & 0.07 \\
26715 & USA 157 & 1 & 5 & Active & 42166 &  & 0.12 & 1 & 1 & 0.07 \\
44012 & ATLAS 5 CENTAUR DEB & 1 & 5 & Unknown & 29065 &  & 0.2 & 0.7 & 0.7 & 0.04 \\
29516 & SINOSAT 2 & 1 & 5 & Inactive & 44374 & 10.0 & 0.11 & 0.3 & 5 & 0.07 \\
5589 & TITAN 3C TRANSTAGE R/B & 1 & 5 & Unknown & 43053 & 2.6 & 0.12 & 0.5 & 1 & 0.07 \\
32387 & RASCOM 1 & 1 & 5 & Inactive & 42508 & 9.0 & 0.12 & 0.4 & 2 & 0.07 \\
30323 & BEIDOU 1D & 1 & 5 & Inactive & 42483 & 20.0 & 0.12 & 0.3 & 6 & 0.07 \\
34705 & IRIDIUM 33 DEB & 1 & 5 & Unknown & 7023 & 0.0 & 13 & 0.003 & 0.2 & 0.0007 \\

\hline
\end{tabular}
\caption{Top ten satellites observed most frequently, and the ten observed just once, according to the cross-identification with the public part of the NORAD satellite catalogue, ordered by the number of matching tracklets. We display the NORAD catalogue ID and name, the number of matching tracklets and alerts, the operational status according to CelesTrak satellite catalogue, semi-major axis of the satellite orbit ($a$), Radar Cross-Section (RCS) also from CelesTrak satellite catalogue, median arc length during ZTF exposure (30 seconds),
and then two light curve period estimators, based on the distances between individual peaks inside alert cutouts ($P_{\mbox{m}}$) and between individual alerts of a tracklet ($P_{\mbox{p}}$), along with the upper limit on the duration of the flashes. Three latter characteristics are derived as discussed in Section~\ref{sec:temporal}.
}
\label{tab:sat_name}
\end{table*}

\subsection{Tracklets without satellite associations}

Among the 6,450 tracklets, 2,609 (40\%) were not matched to objects from the NORAD catalogue. We investigated the possibility of estimating at least some of orbital parameters of the objects producing them directly from the sky positions of individual alerts of a tracklet, but this effort failed. The main reason preventing it is unknown times of appearance of individual alerts inside a single 30-seconds ZTF exposure. Indeed, 
all these alerts happen during the same exposure, and some of them might be missing both due to selection effects (being rejected by the quality cuts, or too faint to be detectable) or due to the object entering or leaving the field of view during mid-exposure, or due to a complex flashing activity pattern that may not span the whole exposure duration. This all make it nearly impossible to reconstruct their relative timing from spatial distances between alerts with any confidence, and without it the orbit determination is sufficiently unstable even for the tracklets with longest arcs on the sky.

We detail in Appendix \ref{sec:tracklet-orbit} our unsuccessful attempts to estimate orbital parameters directly from the tracklets in more details.

\subsection{Temporal properties}
\label{sec:temporal}

\begin{figure}
    \centering
    \centerline{
        \resizebox*{1\linewidth}{!}{\includegraphics{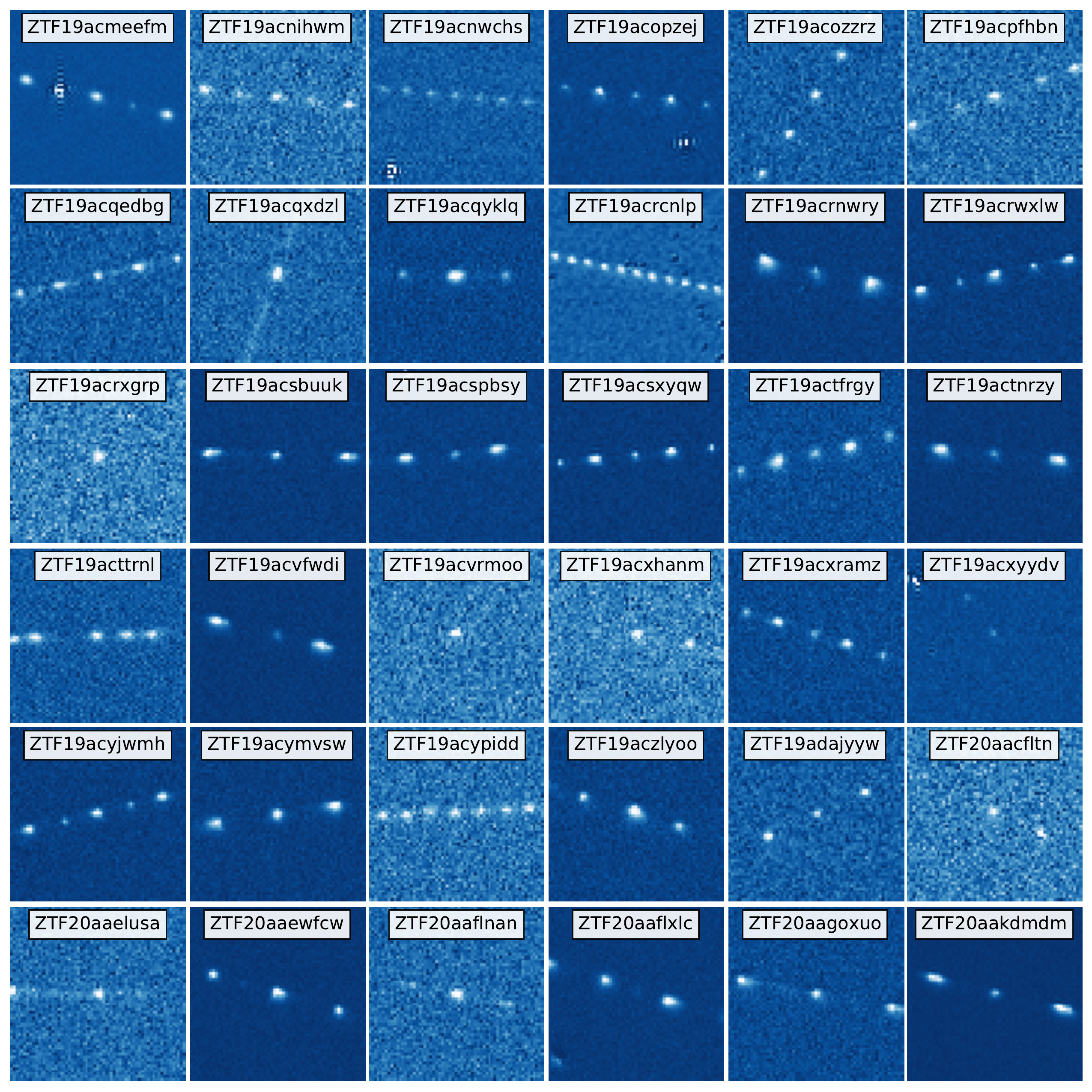}}
    }   
    \caption{Example cutouts of difference images from ZTF alert packets corresponding to a random subset of events associated with tracklets. Large part of them show characteristic multi-peak structure, often symmetric in respect to the ``main'' event, while some fraction is still essentially isolated point sources.}
    \label{fig:cutouts}
\end{figure}

\begin{figure}
    \centering
    \centerline{
        \resizebox*{\linewidth}{!}{\includegraphics{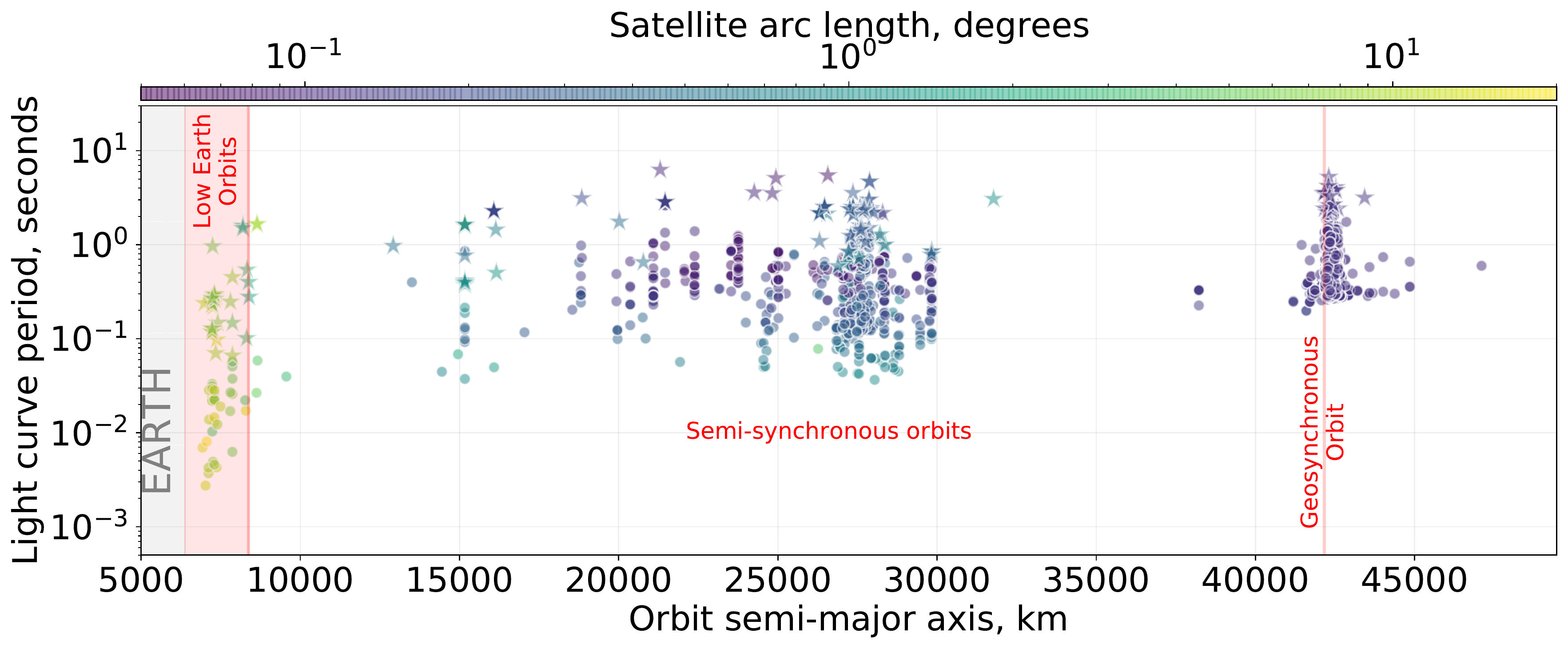}}
    }   
    \caption{Distribution of light curve periods of glinting satellites as a function of their orbit semi-major axes, derived from the spacing of peaks in their cutouts (dots) or spacing of individual events of a tracklet (crosses). Color-coded are the lengths of associated satellite arc on the sky for the ZTF exposure duration of 30 seconds.}
    \label{fig:periods}
\end{figure}

\begin{figure}
    \centering
    \centerline{
        \resizebox*{1\linewidth}{!}{\includegraphics{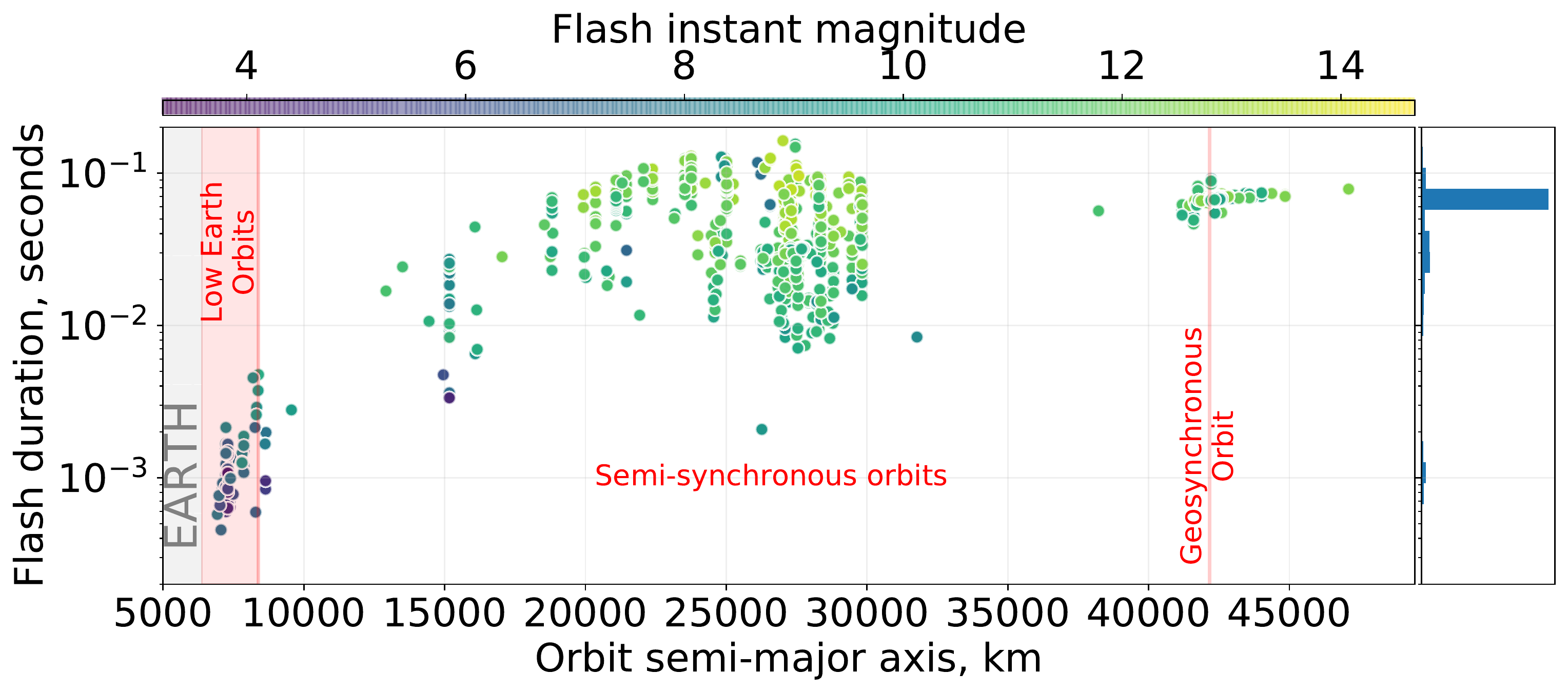}}
    }   
    \caption{Distribution of flash durations of the events associated with known satellites
    as a function of their orbit semi-major axis. Flash duration is derived from the comparison of the length of satellite arc on the sky for the ZTF exposure duration of 30 seconds with the typical FWHM of the images, and thus acts solely as an upper limit for the actual value. Color-coded is the instant brightness of the flashes, measurable in observations with sufficiently high temporal resolution. Smaller panel on the right shows the marginal histogram of durations in the same scale as the main panel.}
    \label{fig:durations}
\end{figure}

Figure~\ref{fig:cutouts} shows how a random subset of events belonging to tracklets look like on cutouts from ZTF difference images, i.e. with all stationary objects (stars, galaxies, etc) removed. Significant fraction of them shows multiple appearances of the transient inside the same cutout (with ~60x60$''$ sizes), implying its rapid periodic variability, supposedly due to fast rotation. In order to characterize them, we performed a simple morphological analysis in the following way.

First, we extracted from \fink\ archive all cutouts corresponding to science images, templates and difference images for all events included in the tracklets detected in Section~\ref{sec:tracklet_detection}. Then we estimated the noise level of science images, and masked all cutout regions where template image has objects above this level. Finally, we normalized the difference image by the noise level of its sky background, estimated as a median absolute deviation (MAD), and performed peak detection in unmasked regions above 2$\sigma$ level using {\sc SEP} code \citep{sep}. We rejected the events where this method does not detect the peak at the cutout center.
This way, we successfully detected peaks for 72,790 (99.2\%) of all events belonging to tracklets, and for 56,461 (77.0\%) of them detected two or more peaks per cutout.
For the latters, we estimated the spacing between individual peaks. Then, knowing both this spacing, and the length of associated satellite arc during the exposure, we may estimate the period of satellite light curve. To make the estimate robust against possible loss of individual peaks in some cutouts, we computed the median value of that period over all cutouts belonging to the same tracklet. The resulting period estimates, one for every tracklet to satellite association, is shown in Figure~\ref{fig:periods}.

For the tracklets where cutouts of individual events do not show multi-peak structure, we may still estimate light curve periodicity using the minimal distance between the tracklet events on the sky. Such estimate will serve as an upper limit on actual light curve period, as not every peak is probably detected, reported in the alert stream and is passing our quality cuts. Figure~\ref{fig:periods} also shows these estimates that are, as expected, systematically longer than the ones derived from the multi-peaked events.

We may also estimate the duration of individual light curve peaks corresponding to the glints by comparing the overall length of associated satellite arc on the sky during the ZTF exposure (30 seconds) with the size of the transients. For the latter, we may assume 1$''$ (half of typical ZTF seeing) to be a good conservative approximation, assuming that significantly larger elongation would lead to the transients failing quality cuts for not being point sources.
Thus, the duration (actually, an upper limit for it) of the flash may be estimated as
\begin{equation}
\tau = 30 \frac{1''}{\mbox{arc length}} \ \mbox{seconds.}
\label{eq:duration}
\end{equation}
On the other hand, instant brightness of the flash peak (the magnitude that would be measured if the temporal resolution of the observations would be high enough to fully resolve it) may be estimated as
\begin{equation}
\mbox{instant magnitude} = \mbox{magpsf} + 2.5\log{\frac{\tau}{30\ \mbox{seconds}}} \ \mbox{,}
\end{equation}
corresponding to the flashes being intrinsically brighter than they appear in ZTF images.
The distribution of the durations of the flashes, as well as their intrinsic brightnesses, estimated this way is shown in Figure~\ref{fig:durations}. 

Visual inspection of the alert cutouts shown in Figure~\ref{fig:cutouts} shows that some of the peaks are in fact elongated -- especially ``secondary'' ones, that not necessarily trigger an alert passing quality cuts. That means that, in addition to shorter and sharper flashes, the light curves may also contain a bit wider ones, with durations 2-3 times longer than the estimates made above.

\section{Discussion}
\label{sec:discussion}

The events associated with tracklets, i.e. definitely caused by satellite glints, constitute up to 11.5\% of all non-repeating alerts in ZTF alert stream that pass our initial quality criteria. This fraction is shown in Figure~\ref{fig:fractions} as a function of event brightness, RealBogus quality score, and distance from the Sun, and is significantly larger for the alerts of moderate (16-17 magnitudes) brightness, better RealBogus quality score and closer to the Sun. 

The absolute rate of tracklet events (number of events per ZTF exposure, see Figure~\ref{fig:rate}) is marginally increasing over time. However, the slope of this increase is not statistically different from the one of overall growth of ZTF alert rate. It is also quite similar to the increase of rate of other classes of events in \fink\ (Solar System candidates, supernova candidates, kilonova candidates, matched alerts in SIMBAD, etc). Thus, we may attribute it to systematic improvements of ZTF observing strategy, data processing and artefact rejection algorithms, and not to a physical increase of satellite glints rate (indeed, it seems rapidly growing satellite mega-constellations are not contributing to this kind of transients).

The fraction of tracklets among all candidate events passing our selection criteria formulated in Section~\ref{sec:initial_selection} shows some signs of an annual variation. However, this effect seems to be dominated by the variation of the rate of candidates, and not tracklet events, and therefore most probably reflects some annual changes in ZTF observing conditions, and not the rate of satellite glints. More data are needed in order to properly characterize the latter.

\subsection{Efficiency of detection}
\label{sec:efficiency}

\begin{figure}
    \centering
     \centerline{
        \resizebox*{1\columnwidth}{!}{\includegraphics{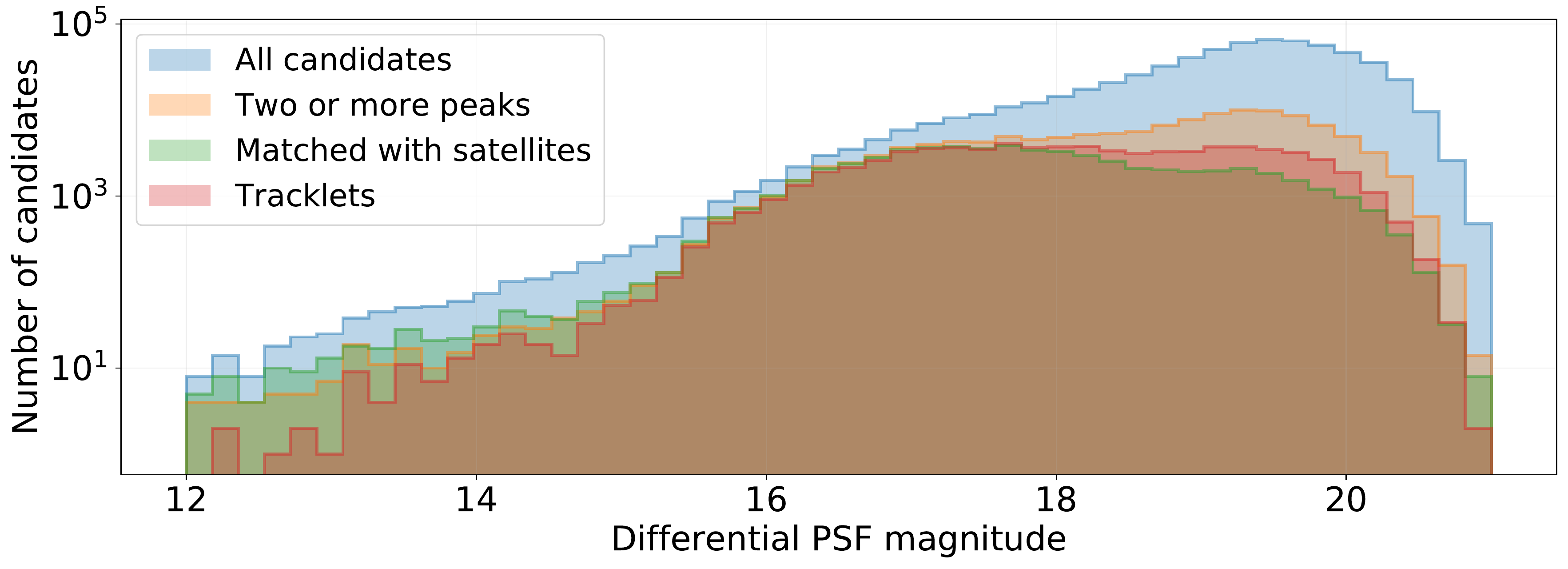}}    
    }
    \centerline{
        \resizebox*{1\columnwidth}{!}{\includegraphics{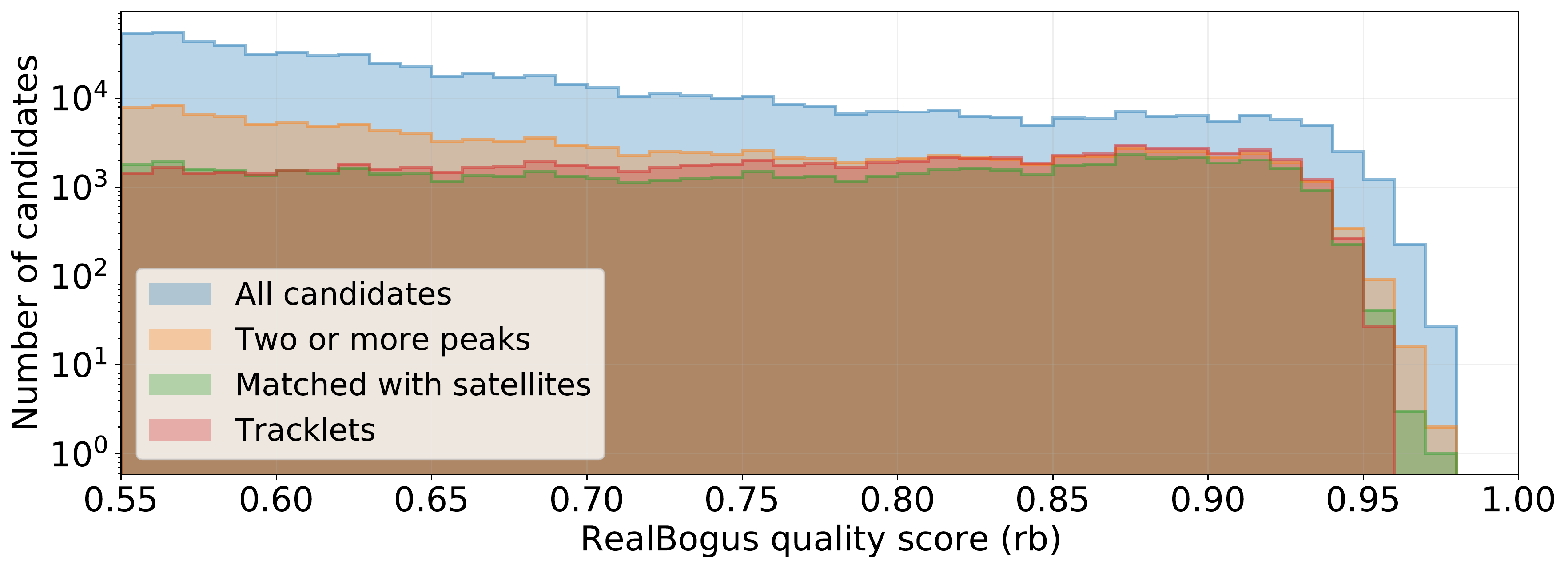}}
    }   
    \centerline{
        \resizebox*{1\columnwidth}{!}{\includegraphics{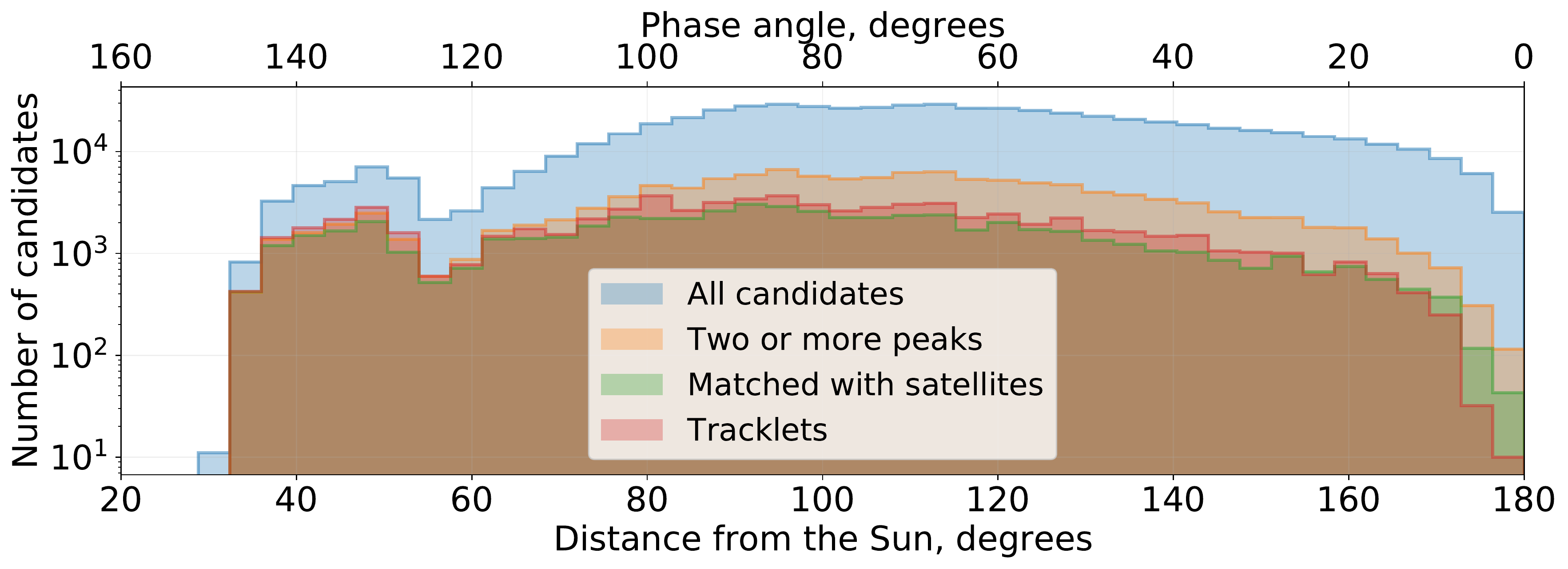}}
    }   
    \caption{Upper panel -- distribution of differential PSF magnitude of all candidate events passing quality cuts as defined in Section~\ref{sec:initial_selection} (blue), as well as subsets where the analysis described in Section~\ref{sec:temporal} detected two or more peaks in difference image cutouts (orange), the ones matched with satellites in Section~\ref{sec:efficiency} (green), and the ones belonging to tracklets detected by the algorithm described in Section~\ref{sec:tracklet_detection} (red).
    Middle panel -- distribution of RealBogus quality score for the same subsets of candidate events.
    Lower panel -- distribution of the angular distance from the Sun (or, alternatively, phase angle defined as a Sun-Object-Observer angle with the vertex at the object) for the same subsets of candidate events.
    }
    \label{fig:fractions}
\end{figure}

We may roughly assess the efficiency of detection of satellite glints among the candidates passing our initial quality cuts by means of geometric (tracklet-based) analysis by performing a ``blind'' matching of every candidate event with the positions of known satellites, and then checking their tracklet status. We did it in a way similar to the one described in Section~\ref{sec:satellites_match}, by propagating orbital elements of all satellites from NORAD public catalogue to the moments of start and end of every ZTF exposure, constructing the great circles corresponding to their arcs, and selecting the candidates from that exposure that are closer than 60$''$ from them. We also rejected the associations where projected distances along the great circle of candidates from first or last points of satellite arc is larger than the length of that arc, thus allowing for some acceptable error in the satellite fly-by time. This way, we acquired satellite associations for 59,421 (9.3\%) candidates. Among them, 44,500 (74.9\%) belong to the tracklets we detected in Section~\ref{sec:tracklet_detection}.

We also applied the morphological analysis of difference image cutouts from Section~\ref{sec:temporal} to all candidate events, and detected two or more peaks in 127,655 (20\%) of them, thus 1.7 times more than amount of events belonging to tracklets. However, as middle panel of Figure~\ref{fig:fractions} shows, this difference is mostly due to the events with lower RealBogus (\texttt{rb}) score, while for \texttt{rb} $>0.8$ the fractions of multi-peak events, the ones blindly matched to satellites, and the ones belonging to tracklets are nearly equal, and contains about 35\% of all candidates. The other source of discrepancy comes from the fact that we require at least 5 alerts to form a tracklet, hence rejecting possible tracks containing fewer alerts (see Fig. \ref{fig:nperframe}).

Therefore we may conclude that the simple tracklet detection algorithm is sufficiently efficient in detecting the events related to the glints of satellites.
It may obviously be complemented by a dedicated alert cutout classification routine, based either on simple morphological criteria used above, or on some full-featured machine learning
algorithms like used in \citet{Carrasco:2020}. That would allow also recovering a (small) subset of alerts that do not pass our tracklet selection criteria. However, we leave this approach for the future, when we will implement it as a part of a full-fledged ``stamp classifier'' for \fink. 

\subsection{Intrinsic brightness of the flashes}
\label{sec:stdmags}

\begin{figure}
    \centering
    \centerline{
        \resizebox*{1\columnwidth}{!}{\includegraphics{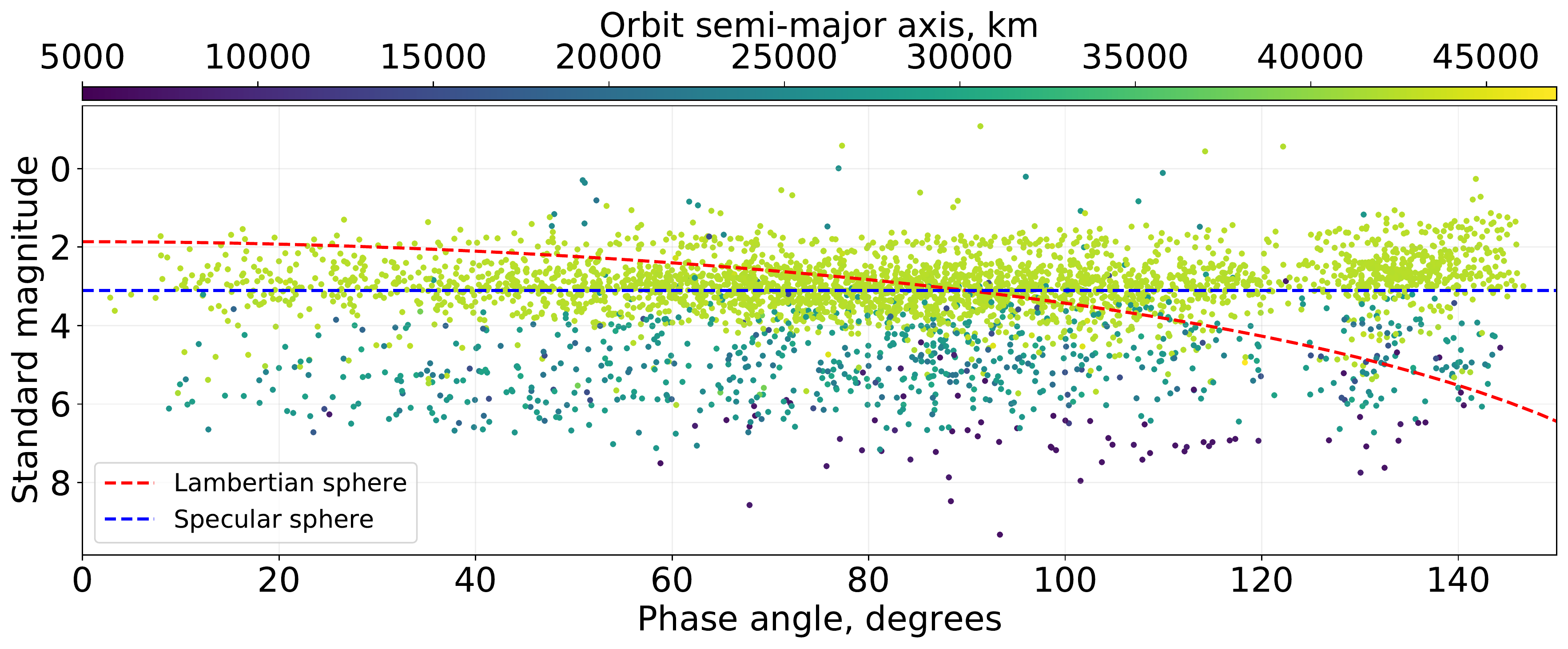}}
    }
    \caption{Standard $g$ band magnitudes of satellite glints as defined in Section~\ref{sec:stdmags}
    as a function of a phase (Sun-satellite-observer) angle.  Overplotted are the functional shapes of expected phase dependence for the lambertian (red dashed line) and specular reflecting (blue dashed line) sphere. The latter seems to be an adequate description for the reflections from some flat mirror-like surfaces on the satellites like solar panels etc.}
    \label{fig:stdmags_phase}
\end{figure}

\begin{figure}
    \centering
    \centerline{
        \resizebox*{1\columnwidth}{!}{\includegraphics{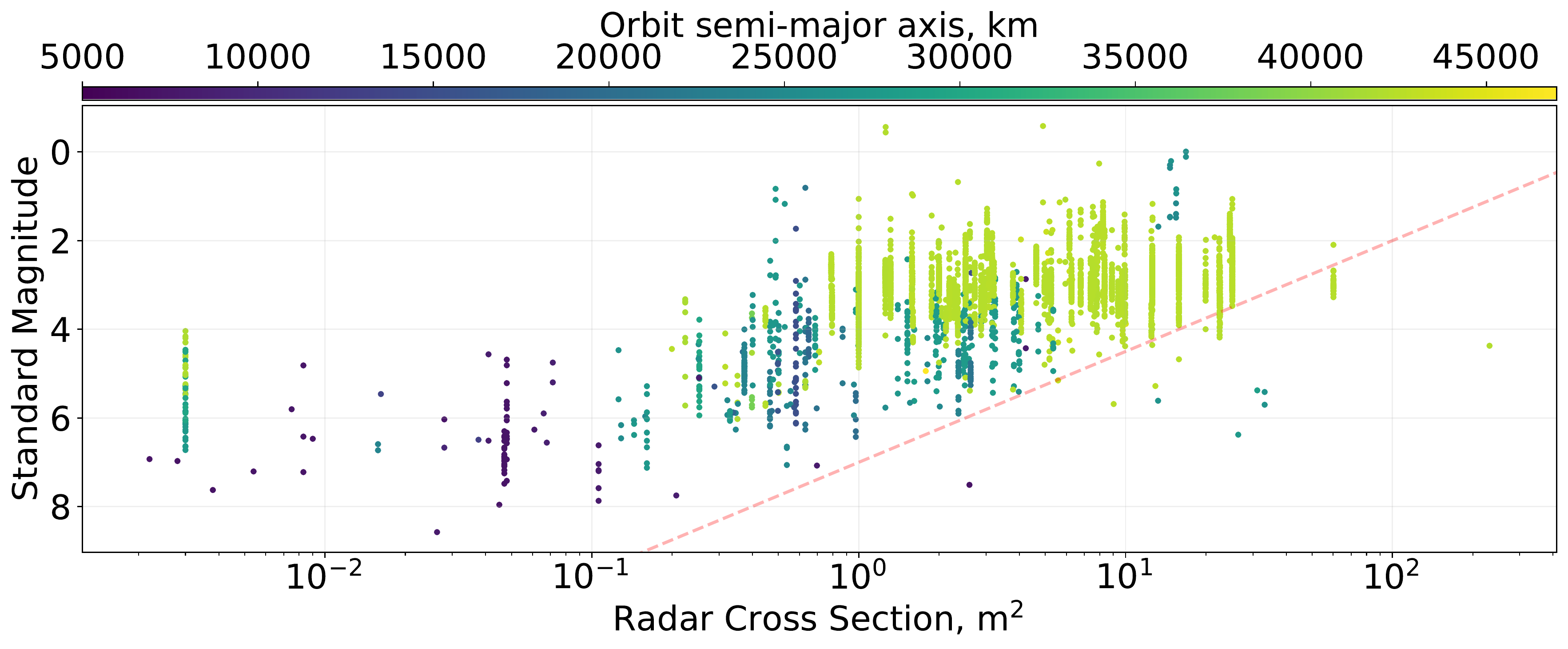}}
    }
    \centerline{
        \resizebox*{1\columnwidth}{!}{\includegraphics{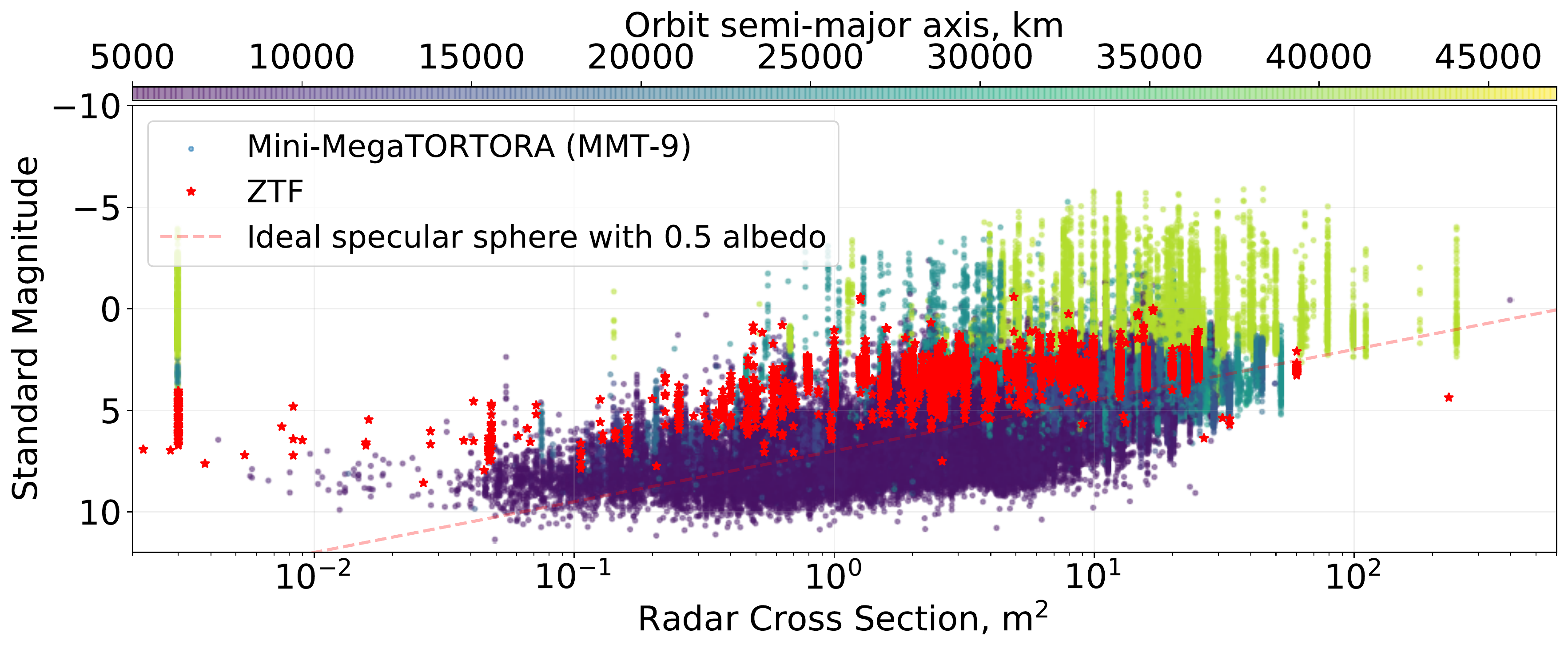}}
    }
    \caption{Upper panel -- standard $g$ band magnitudes of satellite glints as defined in Section~\ref{sec:stdmags} as a function of Radar Cross-Section (RCS) which may be used as a very rough estimation of an overall size of a satellite. Red diagonal line corresponds to the expected brightness of an ideal specular reflecting sphere with 0.5 albedo. The subset of events related to geostationary satellites (a=42,164 km) do not show any dependence of the brightness on RCS, while the ones on lower orbits display some tendency of being fainter for smaller RCS values.
    Lower panel -- the same for  the data from Mini-MegaTORTORA (MMT-9) database of satellite photometry \citep{mmt_satellites}, converted to ZTF $g$ band using Solar spectrum colors. The values measured by ZTF and used in the upper panel are overplotted as red stars.
    }
    \label{fig:stdmags_rcs}
\end{figure}

\begin{figure}
    \centering
    \centerline{
        \resizebox*{1\columnwidth}{!}{\includegraphics{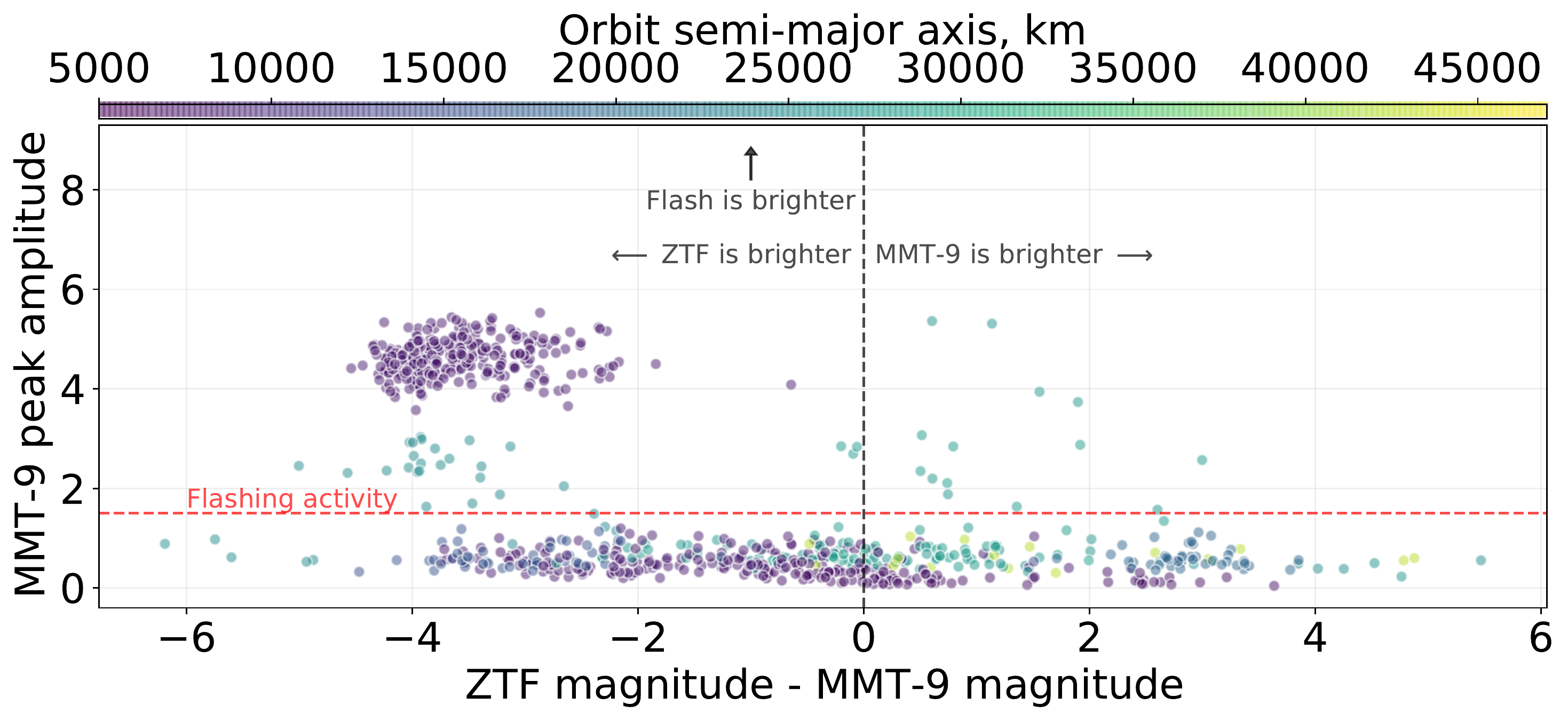}}
    }
    \caption{The amplitude of light curve peaks in Mini-MegaTORTORA (MMT-9) photometric database versus difference of their mean magnitudes as measured by ZTF and mean track magnitudes in Mini-MegaTORTORA data. 
    For Mini-MegaTORTORA data, every individual track is considered individually to better accommodate for different observing conditions; also, the brightness corresponds mostly to the reflection from the main body of the satellite. For ZTF, on the other hand, just a single mean brightness value for every satellite is considered, and it corresponds to the peaks of the glints, i.e. the reflections from some specular reflective surfaces.
    We consider the satellite ``flashing'' in Mini-MegaTORTORA data if the peak amplitude is at least 1.5 magnitudes above the smoothed light curve trends. This division is shown with dashed red horizontal line.}
    \label{fig:stdmags_diff}
\end{figure}

\begin{figure}
    \centering
    \centerline{
        \resizebox*{1\columnwidth}{!}{\includegraphics{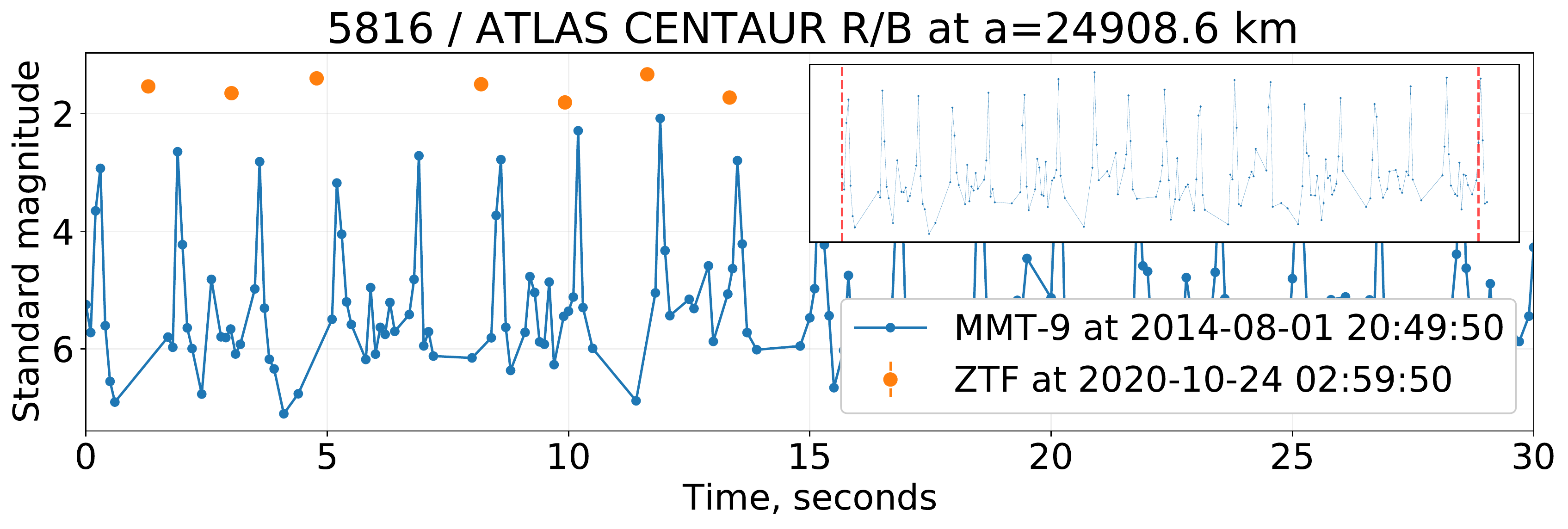}}
    }
    \centerline{
        \resizebox*{1\columnwidth}{!}{\includegraphics{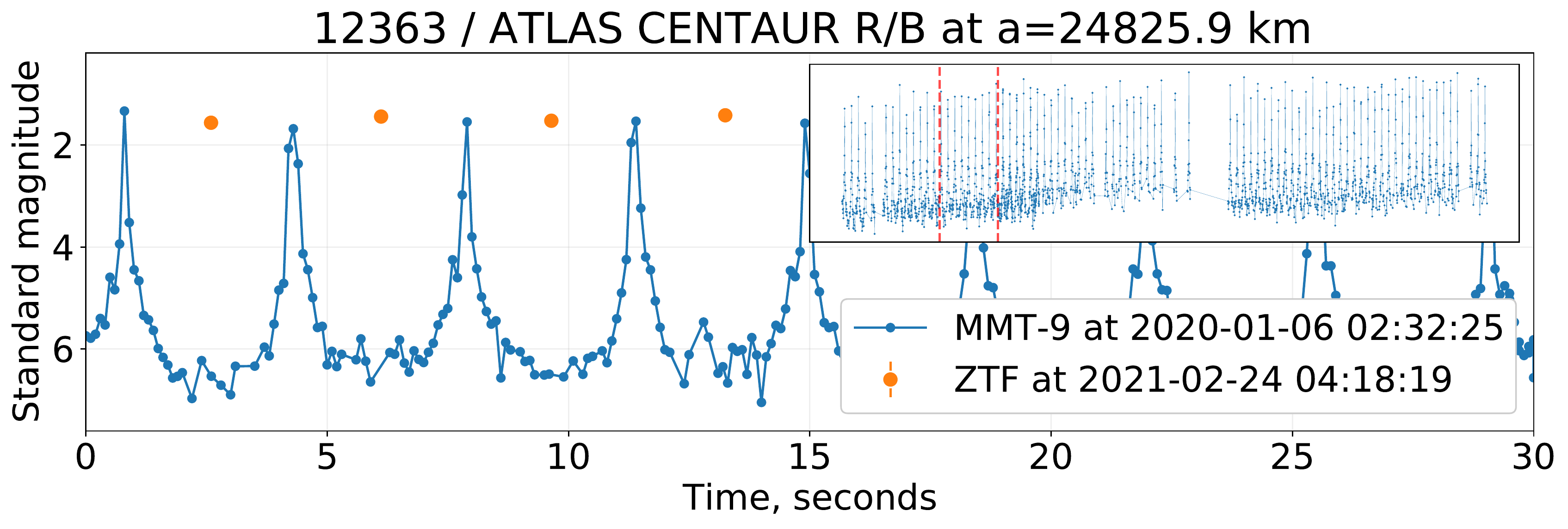}}
    }
    \centerline{
        \resizebox*{1\columnwidth}{!}{\includegraphics{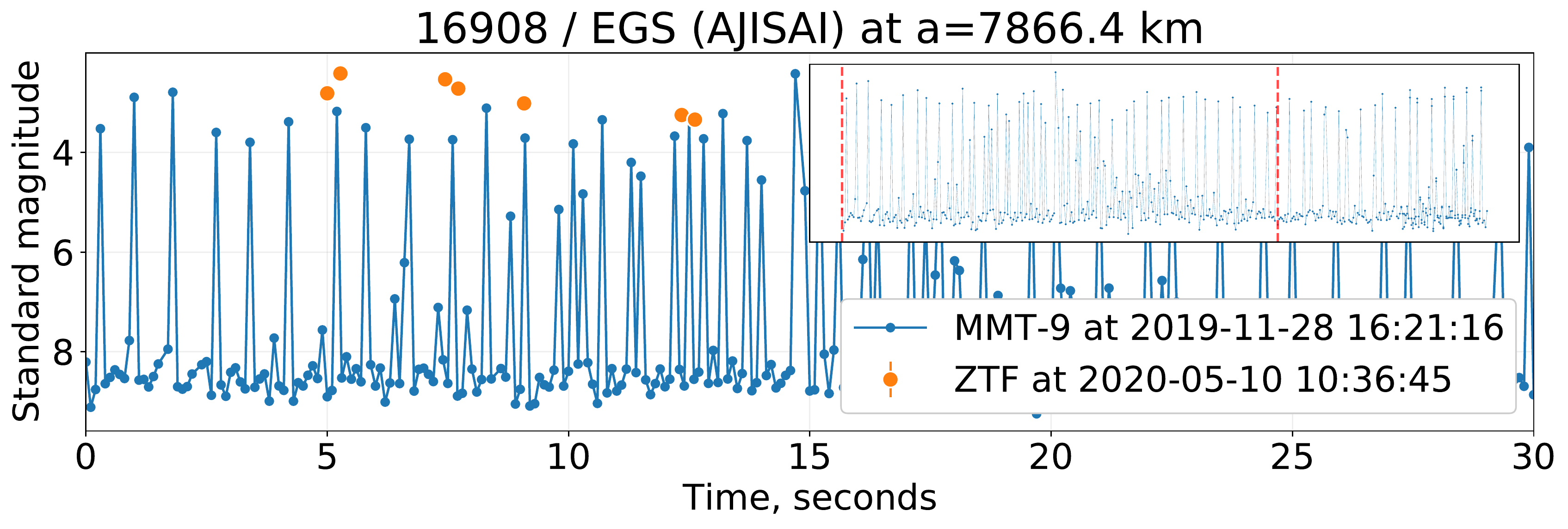}}
    }
    \centerline{
        \resizebox*{1\columnwidth}{!}{\includegraphics{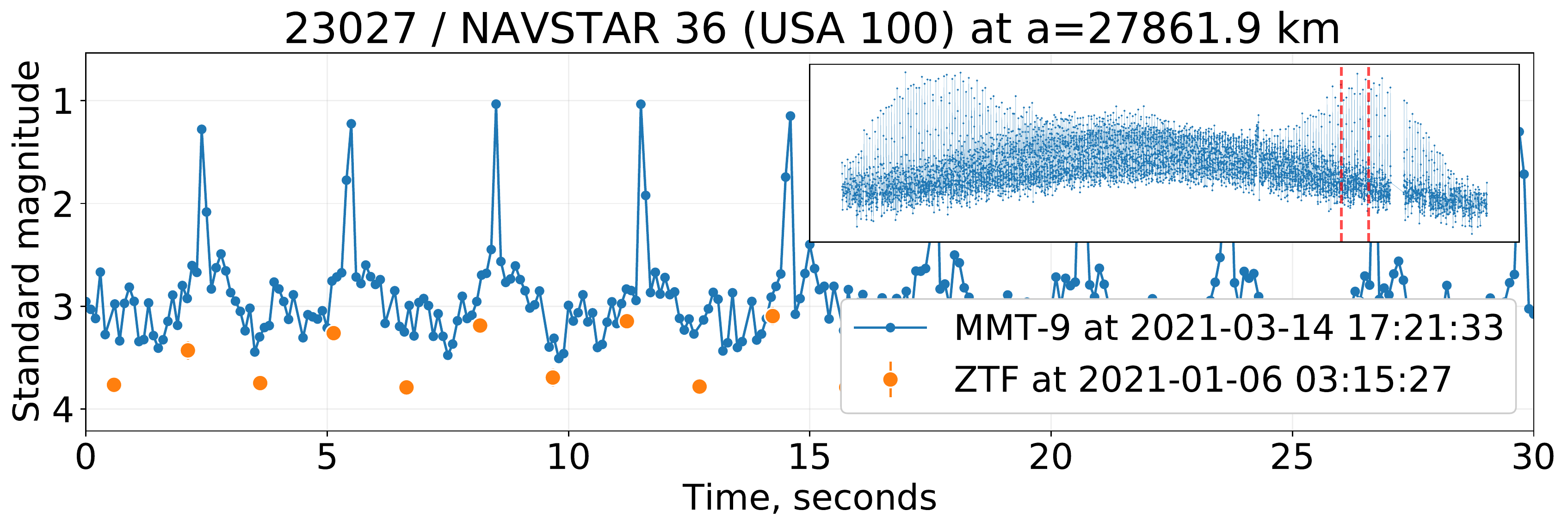}}
    }
    
    \caption{Comparison of the light curves of satellite glints from ZTF data (tracklets) with the typical ones for associated satellites taken from Mini-MegaTORTORA (MMT-9) photometric database. Timings of ZTF light curves have been tentatively reconstructed by comparing angular distances between individual detections in the tracklet and expected associated satellite arc length, and thus have unknown temporal zero point inside the exposure. ZTF and Mini-MegaTORTORA light curves correspond to observations of the same satellite at different times, specified in the plot legends. Inset plot to the right shows an overall light curve of Mini-MegaTORTORA track, with the position of region displayed in the main figure marked with red dashed vertical lines.
    For two first panels corresponding to different Atlas Centaur upper stages, the periods of peaks are in a very good agreement between ZTF and Mini-MegaTORTORA. For the third one, AJISAI satellite, only a few peaks being resolved, probably due to their very tight spacing, overall complex light curve structure and very fast motion of the satellite. Finally, the last one, NavStar 36 satellite, shows complex and highly evolving light curve in Mini-MegaTORTORA data that is brighter than ZTF flashes, having the same main period but lacking fainter interpulses visible in the latters.}
    \label{fig:mmt_lightcurves}
\end{figure}

\begin{figure*}
    \centering
    \centerline{
        \resizebox*{1\columnwidth}{!}{\includegraphics{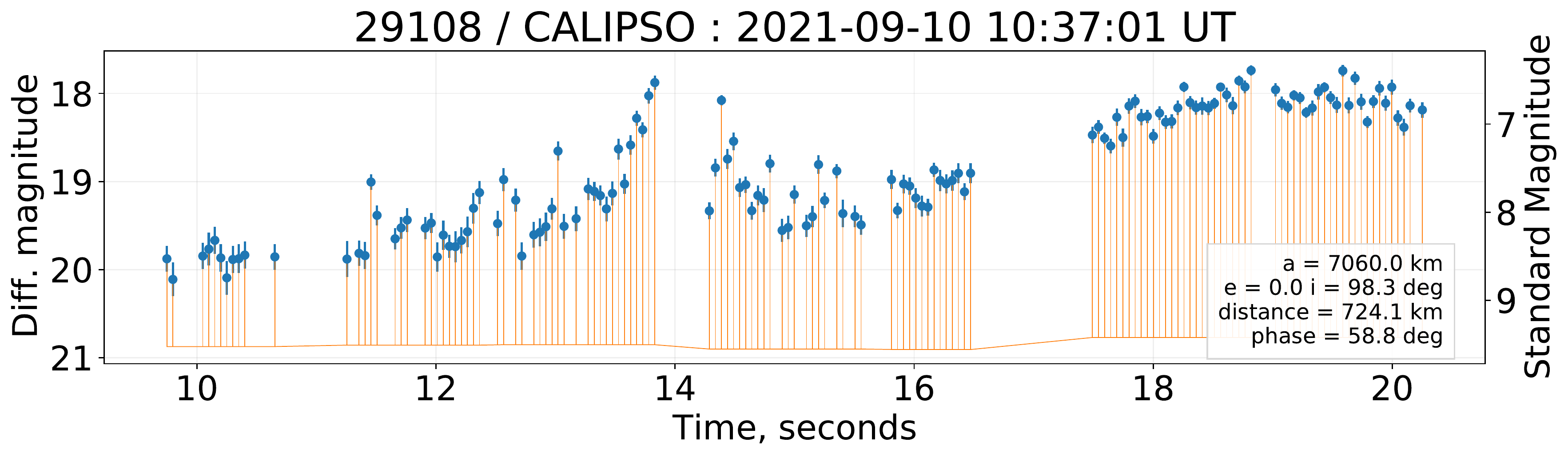}}
        \resizebox*{1\columnwidth}{!}{\includegraphics{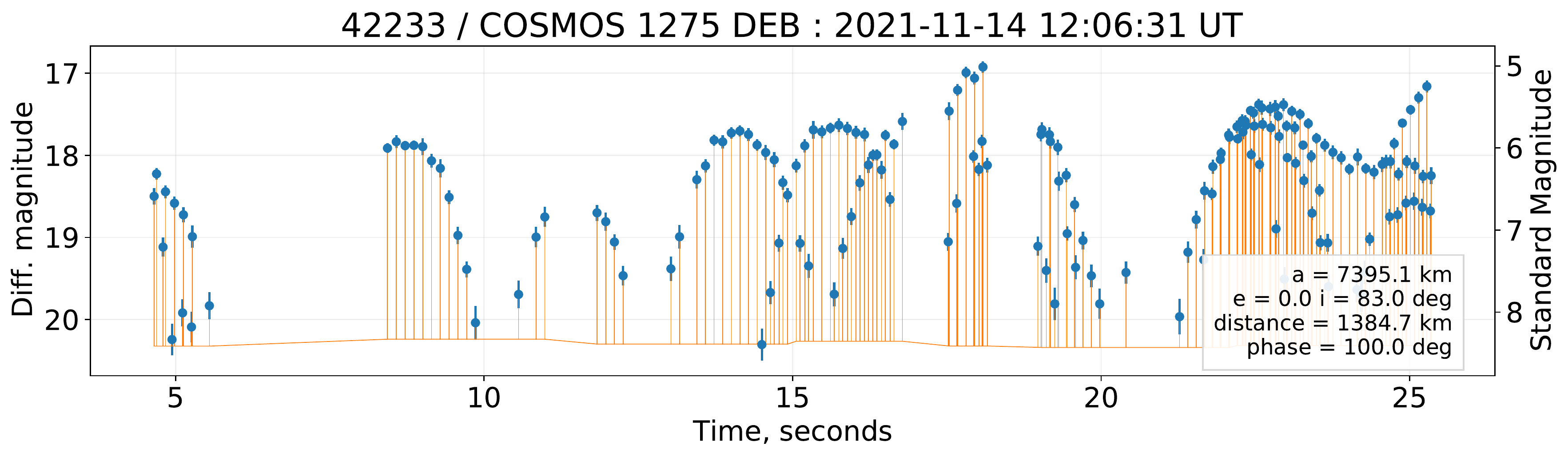}}
    }
    \centerline{
        \resizebox*{1\columnwidth}{!}{\includegraphics{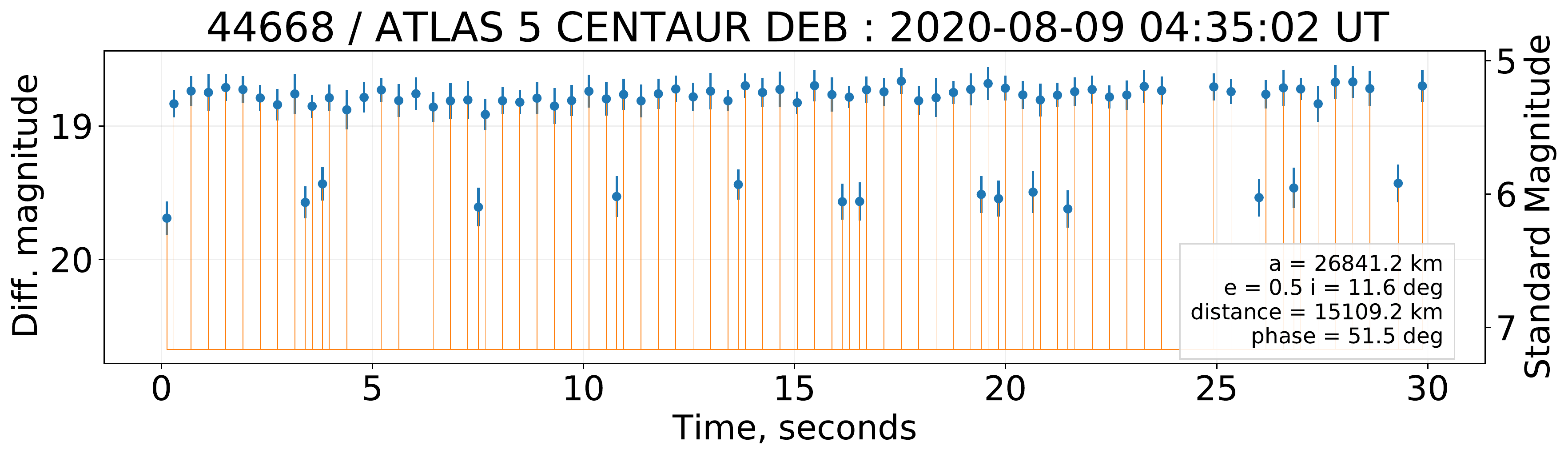}}
        \resizebox*{1\columnwidth}{!}{\includegraphics{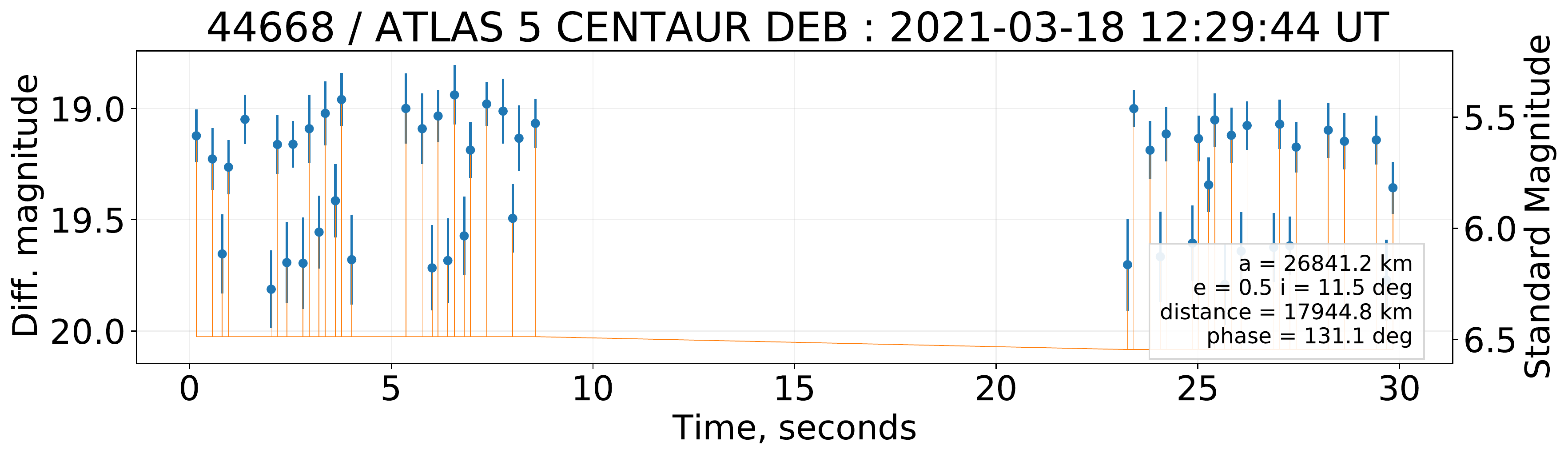}}
    }
    \centerline{
        \resizebox*{1\columnwidth}{!}{\includegraphics{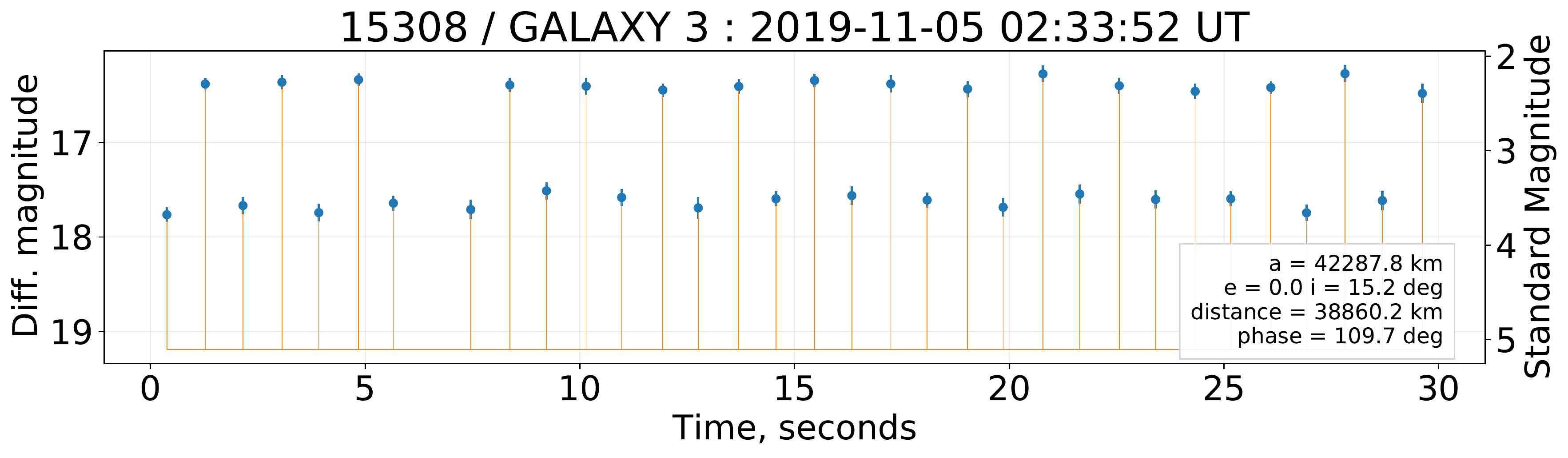}}
        \resizebox*{1\columnwidth}{!}{\includegraphics{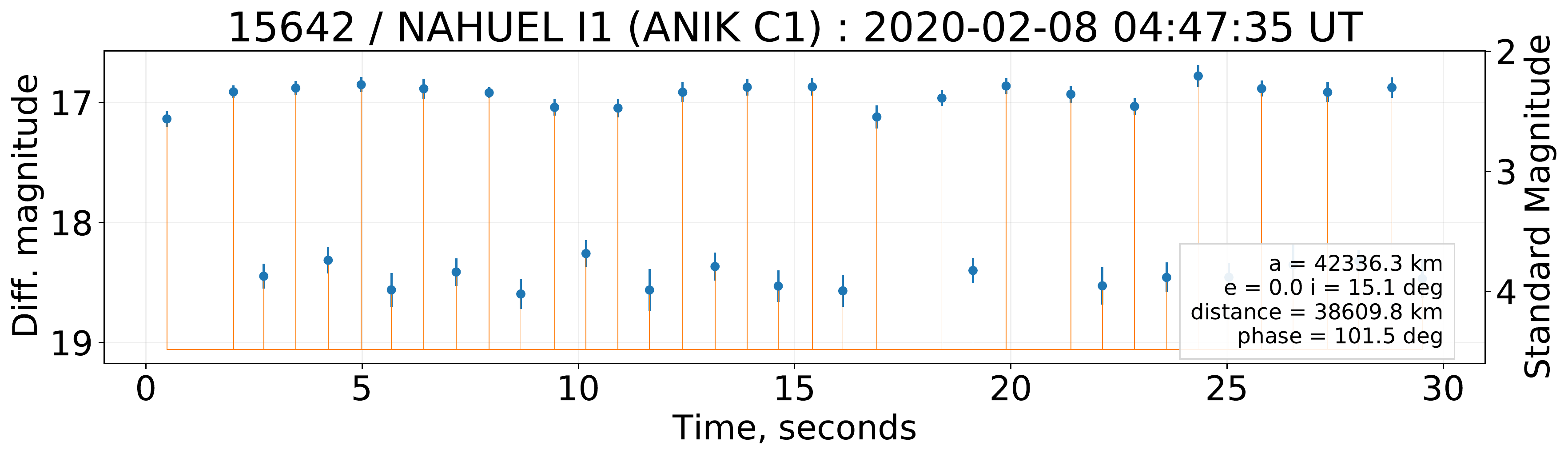}}
    }
    \centerline{
        \resizebox*{1\columnwidth}{!}{\includegraphics{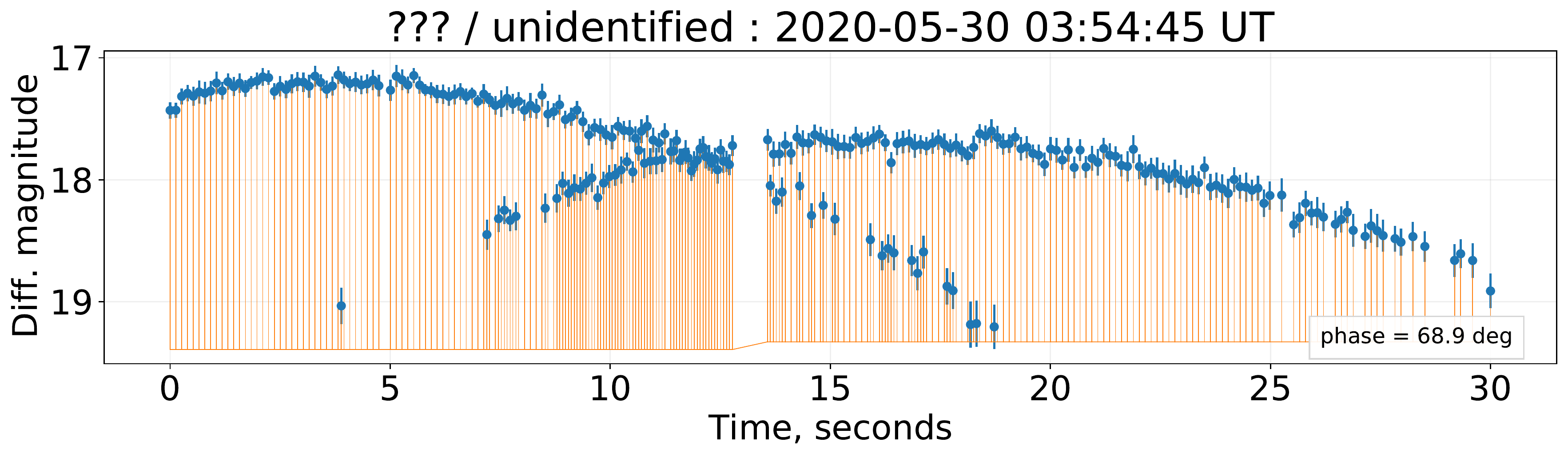}}
        \resizebox*{1\columnwidth}{!}{\includegraphics{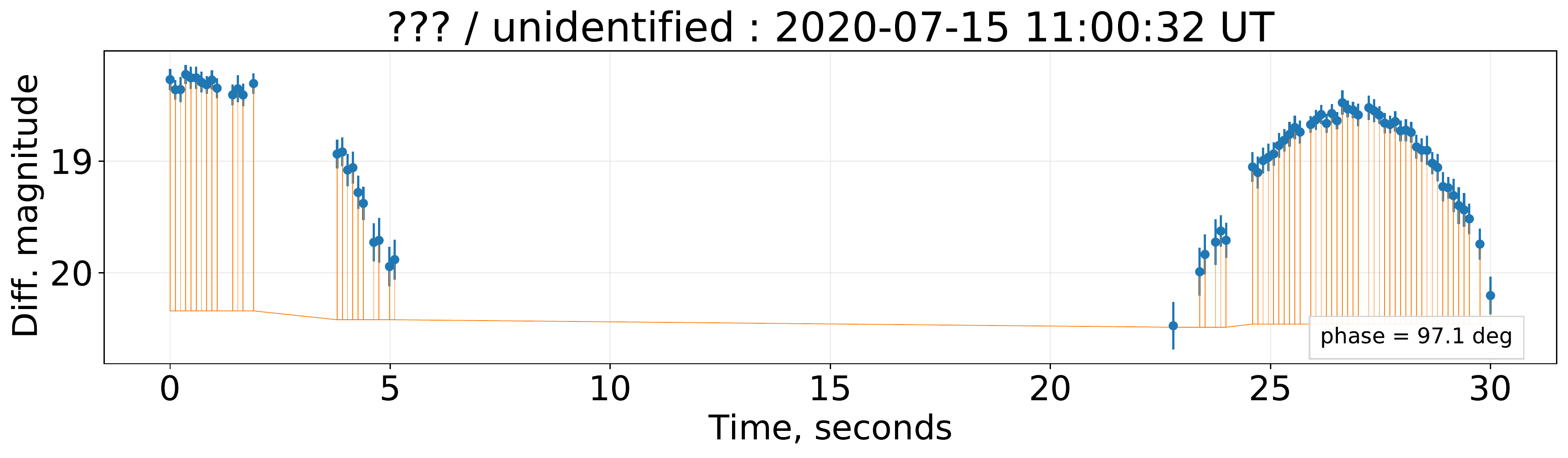}}
    }
    \caption{Light curves of the tracklets as seen by ZTF. Rows from top to bottom show examples of the events associated with satellites on low-altitude orbits, medium orbits, geostationary orbits, and not associated with catalogued satellites, correspondingly. For the events associated with satellites the plots also show their orbital parameters and standard magnitudes. The time scale (horizontal axis) is reconstructed using the relative spacing between the tracklet peaks on the sky and the length of satellite track arc (for events associated with satellites) or empirical tracklet arc length (for unassociated ones). Horizontal orange line corresponds to the ZTF differential imaging detection limit as reported for the alerts (``diffmaglim'' field in AVRO alert schema), and thus may be considered as an upper limit for ``quiescent'' brightness component of the satellite between the glints.}
    \label{fig:lcs}
\end{figure*}

\begin{figure}
    \centering
    \centerline{
        \resizebox*{1\columnwidth}{!}{\includegraphics{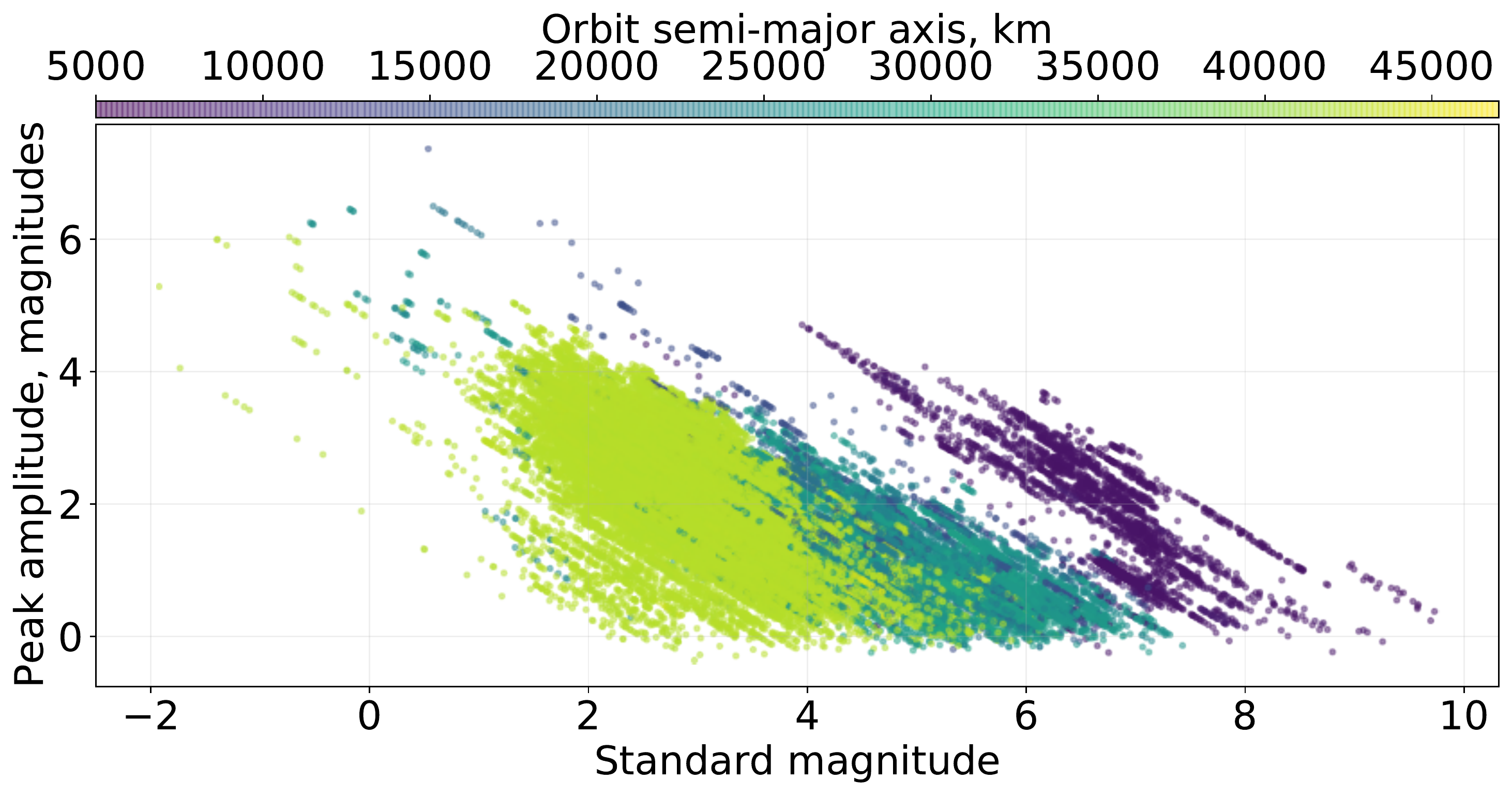}}
    }
    \caption{The amplitudes of light curve peaks in ZTF data for the events associated with known satellites as a function of their standard magnitudes and orbital parameters. The amplitudes are estimated from differential PSF magnitudes and differential detection limits of individual alerts, and are thus lower limits for actual flash amplitudes.}
    \label{fig:stdmags_amplitudes}
\end{figure}

In order to better characterize the intrinsic brightness of flashes from the population of glinting satellites we are seeing in ZTF data, we may compute ``standard magnitudes'' for them by normalizing their instant brightness, as defined in Section~\ref{sec:temporal}, to a fixed observing distance of 1000 km. This way we may both negate the effect of limited temporal resolution of the ZTF observations, and the difference of observing distances between both different satellites and different observations of the same object at different positions of its orbit. We define ``standard magnitude'' using the following prescription for the event magnitude measured on ZTF image:
\begin{equation}
\mbox{magpsf} = \mbox{stdmag} + F(\phi) + 5\log{(\frac{d}{1000\ \mbox{km}})} - 2.5\log{(\frac{\tau}{30\ \mbox{s}})} \ \mbox{,}
\end{equation}

where $d$ is the distance to the satellite, $\tau$ is a true duration of the flash (we will use the upper limit from Eq.~\ref{eq:duration} instead), and $F(\phi)$ is the function describing phase ($\phi$, defined as a Sun-satellite-observer angle) dependence of the apparent brightness due to varying solar illumination of the satellite . The latter is usually defined in such a way that this term is zero at $\phi=90$ degrees (thus, standard magnitude corresponds to apparent magnitude of a satellite at 1000 km distance and 90 degrees phase angle).
For a diffuse Lambertian sphere, the function is 
\begin{equation}
F(\phi) = -2.5\log{[\sin{\phi} + (\pi-\phi)\cos{\phi}]} \ \mbox{,}
\end{equation}
while e.g for a specular (mirror-like) reflective sphere directing the solar beam towards the observer, it does not depend on the phase 
\begin{equation}
F(\phi) = 0 \ \mbox{.}
\end{equation}
For the ease of comparing the magnitudes measured in different filters we will also convert all the measurements to ZTF $g$ bandpass assuming that the spectrum of events is essentially solar, and thus its color is $g-r=0.46$. All the estimations of standard magnitudes below will therefore correspond to ZTF $g$ band.

Figure~\ref{fig:stdmags_phase} shows the dependence of standard magnitudes for the events associated with known satellites, computed without phase-dependent term, on the phase angle. It is clear that the phase dependence is essentially absent, which is consistent with the glints being produced by specular (mirror-like) reflections from some flat surfaces on the satellites. Thus, the specular sphere approximation seems to be an adequate description for the brightness of the peaks of the flashes, and we will use this approximation for standard magnitudes below.

Upper panel of Figure~\ref{fig:stdmags_rcs} shows the dependence of standard magnitudes on the Radar Cross-Sections (RCS) of the associated satellites, which may act as a very rough estimate for an overall size of the satellite. We may clearly see that for geostationary satellites the brightness is nearly independent of RCS and exceeds an expected brightness for an ideal specular sphere. That may be related to the fact that specular plane should reflect much more light in the direction of a a sunbeam than an uniform specular sphere; moreover, published RCS values most probably correspond to an angle-averaged satellite cross-section which may be significantly smaller than reflecting area of e.g. solar panels.
 
Now that we have standard magnitudes of the glints, independent from specific observing conditions on ZTF, we may compare their brightness with the data from the largest public database\footnote{Available online at \url{http://mmt.favor2.info/satellites} and contains photometric data for more than 360,000 
individual tracks of more than 9,500 satellites observed with 10 frames per second sampling rate.} of satellite photometry \citep{mmt_satellites} that contains the results of satellite brightness measurements from Mini-MegaTORTORA \citep{karpov_2019} -- a set of nine wide-field cameras operating in white light at 10 frames per second and calibrated to Johnson V filter zero point.
We extracted a small random subset of photometric measurements from that database and converted them to ZTF $g$ band using solar $g-V=0.27$ color. We also applied diffuse Lambertian sphere phase angle correction in order to compensate significant phase angle dependence of apparent brightness (that corresponds to most of the measurements being from the primary bodies of the satellites, not from specular surfaces like solar panels). We did not apply any correction for the temporal resolution, as exposures of Mini-MegaTORTORA cameras are sufficiently short to resolve the peaks of most of the flashes. The standard magnitudes for all these measurements are shown in lower panel of Figure~\ref{fig:stdmags_rcs}. The brightness of satellite glints seen by ZTF is on the brighter end of an overall cloud corresponding to satellites on lower orbits, while for the geostationary ones they are systematically fainter than the Mini-MegaTORTORA measurements. The latter is due to limited ($V_{\mbox{lim}}\approx11$ mag) sensitivity of Mini-MegaTORTORA observations that is often not enough to follow the objects on these orbits that have $\approx$8 magnitudes distance correction factor. Moreover, the Mini-MegaTORTORA database contains only brightness measurements for the satellites clearly visible for longer intervals of time, while isolated flashes like the ones we see in ZTF data are not catalogued there, despite still being detected \citep{mmt_flashes}.

\subsection{Light curves}
\label{sec:lightcurves}

We also extracted from Mini-MegaTORTORA database the complete light curves available for 59  satellites we identified in ZTF data. For them, 345 individual tracks (i.e. sets of consecutive observations corresponding to a single fly-by of a satellite) of 11 different satellites show significant (with more than 1.5 magnitudes amplitude) flashing activity on a timescale shorter than a second. Figure~\ref{fig:stdmags_diff} shows the amplitude of this activity (defined as a maximal difference between track light curve peaks and a ``smooth'' component constructed by applying a second-order Savicky-Golay filter with 1.1 seconds window to the light curve) as a function of the difference of mean magnitudes in ZTF and Mini-MegaTORTORA measurements, both converted to standard magnitudes as described above. The figure clearly shows that typical amplitudes of the flashes of those satellites are significant, at least 4 to 6 magnitudes in the majority of cases. 
Also, for most low-orbit satellites mean brightness in Mini-MegaTORTORA data, that mainly corresponds to a ``quiescent'' parts of the light curve, is significantly fainter than the brightness of ZTF measurements that corresponding solely to light curve peaks. On the other hand, much more sparse data on higher-orbit and geostationary satellites correspond to their much brighter overall appearance, probably due to different illumination conditions of the main satellite bodies. Due to limited sensitivity of Mini-MegaTORTORA cameras, fainter tracks of geostationary satellites cannot be observed there. Also, no geostationary satellite shows any apparent flashing activity in Mini-MegaTORTORA data, most probably due to being outshined by a much brighter ``quiescent'' reflected light.

In general, the shapes and peak brightness of light curves of these satellites in Mini-MegaTORTORA data shown in  Figure~\ref{fig:mmt_lightcurves} are quite consistent with the parameters of ZTF flashes, including the durations of the peaks (they are not resolved with 0.1 seconds sampling of Mini-MegaTORTORA light curves, which is consistent with the estimations made in above and shown in Figure~\ref{fig:durations}). Figure~\ref{fig:lcs} also shows some examples of reconstructed light curves for longer ZTF tracklets, with the brightness of ``quiescent'' segments of the light curve estimated through the ZTF differential imaging detection limit (thus, it is actually an upper limit for these parts of the actual light curve), and light curve peaks assumed to be infinitely narrow. The objects on lower orbits seem to display the most chaotic behaviour, especially the remnants of COSMOS~1275 explosion that show very complex patterns consisting of multiple reflection components. On higher orbits the behaviour is generally more stable, with just a single or double periodic peaks of slowly drifting amplitudes. Unidentified events show also complex light curve structures suggesting that they are smaller low-altitude objects with probably too fast evolving orbits that prevents their association with satellite catalogues.

\subsection{Impact on upcoming LSST survey}

Vera C. Rubin is currently nearing its completion, and the Legacy Survey of Space and Time (LSST) that will soon start on its Simonyi Survey Telescope will obviously also be impacted by satellite glints. We may estimate the impact as follows.

Expected seeing at Rubin observatory (FWHM of 0.8$''$, properly sampled with 0.2$''$ pixels) is more than 2 times better than on ZTF, thus for the flashes to be still detectable as point sources they have to be at least two times shorter than upper limits shown in Figure~\ref{fig:durations}. As we do not know actual durations of the glints, except for that they are shorter than 0.1 seconds (see discussion of Mini-MegaTORTORA data in Section~\ref{sec:lightcurves}), and as the distribution shown in Figure~\ref{fig:durations} is quite wide, we may assume that this condition will still hold for LSST. The effective duration of individual LSST Deep-Wide-Fast visits \footnote{The visit will contain two consecutive 15 seconds exposures, and image differencing and transient detection will be performed solely on the coadds of those, thus increasing effective visit exposure to 30 seconds.} \citep{lsst} will be the same as ZTF (30 seconds) exposures, so the flashes of the same durations will appear there with the same apparent magnitudes as in ZTF.
On the other hand, finer
spatial resolution and better observing conditions at Vera C. Rubin observatory will lead to $2''/0.8''=2.5$ times shorter effective exposure (defined as a crossing time for an image resolution element, e.g. FWHM) for moving targets, thus making the satellite trails $\approx$1 magnitude fainter when viewed by Rubin Observatory. The detection limit will be approximately 4 magnitudes better in LSST (at least $g=24.5$ for realistic observing conditions, compared to nominal $g=20.5$ detection limit for the same exposure time), thus the satellite trails in LSST will be detectable up to 3 magnitudes deeper.

We do not know actual ``quiescent'' brightness of the majority of satellites producing the flashes we are detecting. However, from Figure~\ref{fig:stdmags_diff} (also Figure~\ref{fig:mmt_lightcurves} and, under some assumptions, lower panel of Figure~\ref{fig:stdmags_rcs} too) we may see that they are typically up to 5 magnitudes fainter than peak brightness of the flashes. Comparing that with the distribution of peak amplitudes of flashes above ZTF detection limit in Figure~\ref{fig:stdmags_amplitudes} which broadly spans from 0 to 4 magnitudes, we may expect that for significant part of these events the ``quiescent'' part of the trail will still be below the LSST detection limit. As the amount of glints from every individual satellite during the exposure will be the same, they will also be detectable by the algorithm discussed in Section~\ref{sec:tracklet_detection}.

Thus, we may expect an appearance of significant amount of very bright flashes from the satellites we are routinely detecting in ZTF data, along with possibly much larger amount of fainter ones due to smaller satellites and debris objects that are below ZTF detection limit. 

It is difficult to predict the statistics of satellites and debris detectable at deeper magnitudes due to general lack of reliable photometric information on them. However, e.g recent study of fainter debris on geostationary orbits by \citet{debriswatch} revealed a large number of uncatalogued objects with significant amount of ``tumbling'' ones, i.e. showing clear signs of rotational and flashing activity, among them. Authors also observed a clearly bimodal distribution of their brightness, with peaks corresponding, using the same notation as in Section~\ref{sec:stdmags} above, to standard magnitudes of 4 (mostly catalogued objects) and 14 (uncatalogued ones). If this brightness corresponds to an overall ``quiescent'' light curve, and assuming that the flashes may have an amplitude of up to 5-6 magnitudes above it, the first one is sufficiently brighter than ZTF tracklets we detected (and should mostly produce clearly visible trails in both ZTF and LSST), while the second peak is exactly on the level of being barely detectable by LSST. Thus, the majority of debris located between these peaks will be easily apparent there. The same arguments should be also correct for similar uncatalogued debris components on lower orbits as well, as, while they are intrinsically brighter for the same size than on geostationary orbit, higher angular velocity makes them harder to detect and study by typical debris survey experiments. Thus, we may conclude that this fainter debris population will be significant contributor to the budget of rapid optical transients in Vera C. Rubin Legacy Survey of Space and Time (LSST).

\section{Conclusion}
\label{sec:summary}

We have developed a simple methodology for locating the events caused by satellite reflection glints, but looking mostly as genuine point source transients, from the ZTF data stream, and implemented a code that routinely extracts and marks them in the framework of \fink\ alert broker. The routine is based on simple geometric considerations, and it successfully detects the majority of events associated with known satellites or showing distinct multi-peak patterns in their cutouts. In general, the alerts belonging to the tracklets, and thus produced by glinting satellites, account for up to 11.5\% of all single-frame events (and up to 30\% of the ones with RealBogus quality score better than 0.8) after applying some quite basic quality cuts not inherent for near-Earth objects but also e.g. for rapid transients of properly astrophysical origin.

The glinting satellites occupy all possible orbits around the Earth, from low to medium to geosynchronous ones, and the durations of the observed flashes are as short as $10^{-1}$--$10^{-3}$ seconds with instant brightness of $g$=4--14 magnitudes. Their light curves show in general quite complex evolution of glinting activity over time, consistent with the light curves of glinting satellites in photometric database of Mini-MegaTORTORA project.

The relatively short arcs of the detected tracklets on the sky, and the absence of intrinsic timing information for individual tracklet alerts within the exposure make it impossible to directly assess the orbits for the tracklets that are not associated with known satellites. However, the amount of data may still allow for some statistical inference on the population of these objects and their properties, that may be valuable for the study of debris populations on various orbits around the Earth.

In ZTF alert stream, the amount of glinting events detectable with the proposed routine is 0.6 per exposure and 140 per night. Assuming that the proportion of objects orbiting around the Earth increases with the decrease of their size, deeper surveys such as the Vera C. Rubin LSST will detect much more events of this kind, and thus are in need for this sort of routines to properly handle and characterize them.

\section*{Software packages used}

\fink\ makes extensive use of several libraries and frameworks among which projects from the Apache Software Foundation\footnote{\url{https://apache.org/}} (Apache Hadoop, Apache HBase, Apache Kafka, Apache Spark), AstroPy, NumPy, matplotlib, Pandas, scikit-learn \citep{pandas:2010,numpy:2011,astropy:2018,astropy:2013,Hunter:2007,scikit-learn:2011}. The research also makes use of SEP \citep{sep} and Orekit \citep{orekit}.

\section*{Data availability}

All tracklet data from \fink\ since November 2019 is publicly available, as is all \fink\ processed data, through \fink\ Science Portal at \url{https://fink-portal.org}. The procedure to access the tracklet data is detailed in Appendix \ref{sec:science-portal}. The results of association of tracklets with satellites are available upon reasonable request to the authors.

\section*{Acknowledgements}

This work was developed within the \fink\ community and made use of the \fink\ community broker resources. \fink\ is supported by LSST-France and CNRS/IN2P3. We acknowledge the support from the VirtualData cloud at Universit{\'e} Paris-Saclay which provided the computing resources.
SK acknowledges support from the European Structural and Investment Fund and the Czech Ministry of Education, Youth and Sports (Project CoGraDS-CZ.02.1.01/0.0/0.0/15003/0000437).
This research uses data from Mini-MegaTORTORA, which belongs to Kazan Federal University, and whose operation is performed according to the Russian Government Program of Competitive Growth of Kazan Federal University. 
This work was partially supported within the framework of the government contract of the Special Astrophysical Observatory of the Russian Academy of Sciences in its part entitled ``Conducting Fundamental Research''.

\small
\bibliographystyle{mnras}
\bibliography{ref}



\appendix

\section{Orbit determination for tracklets}\label{sec:tracklet-orbit}

\begin{figure}
    \includegraphics[width=1\linewidth]{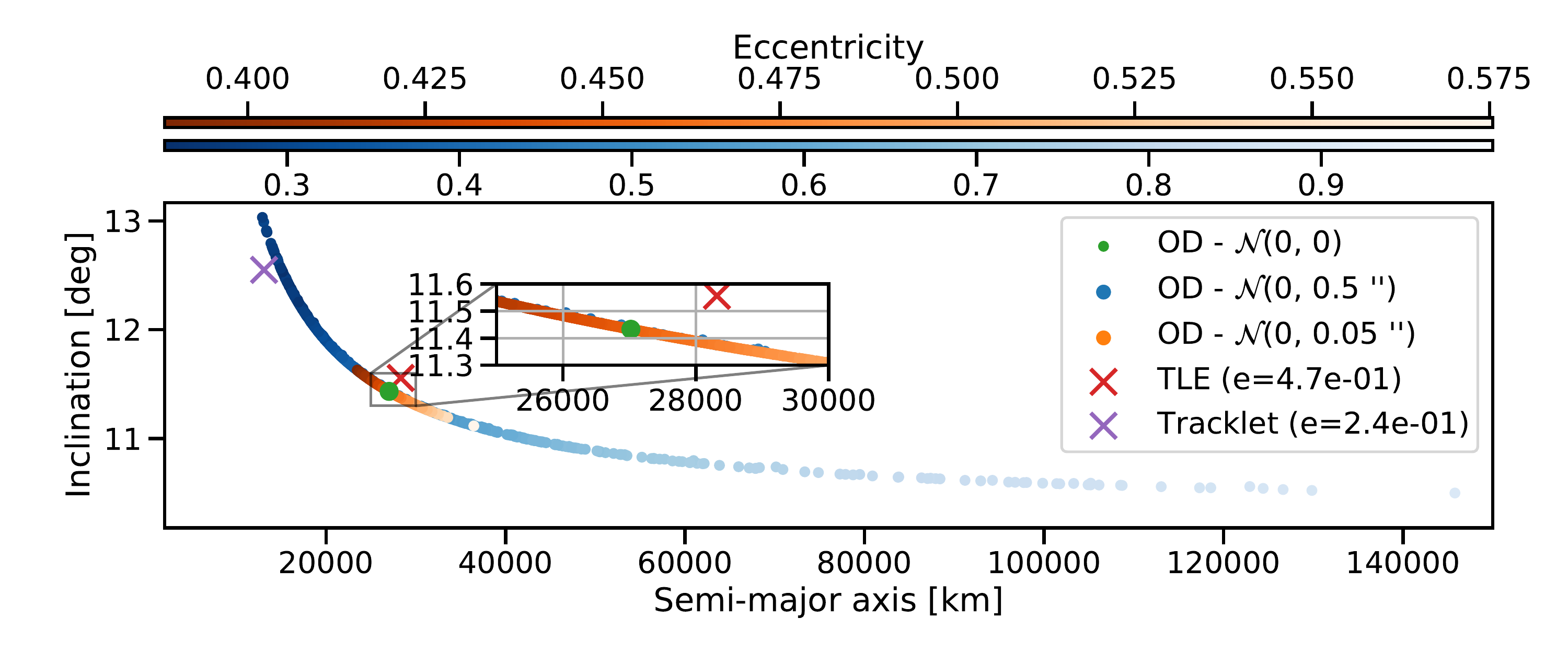}
    \includegraphics[width=1\linewidth]{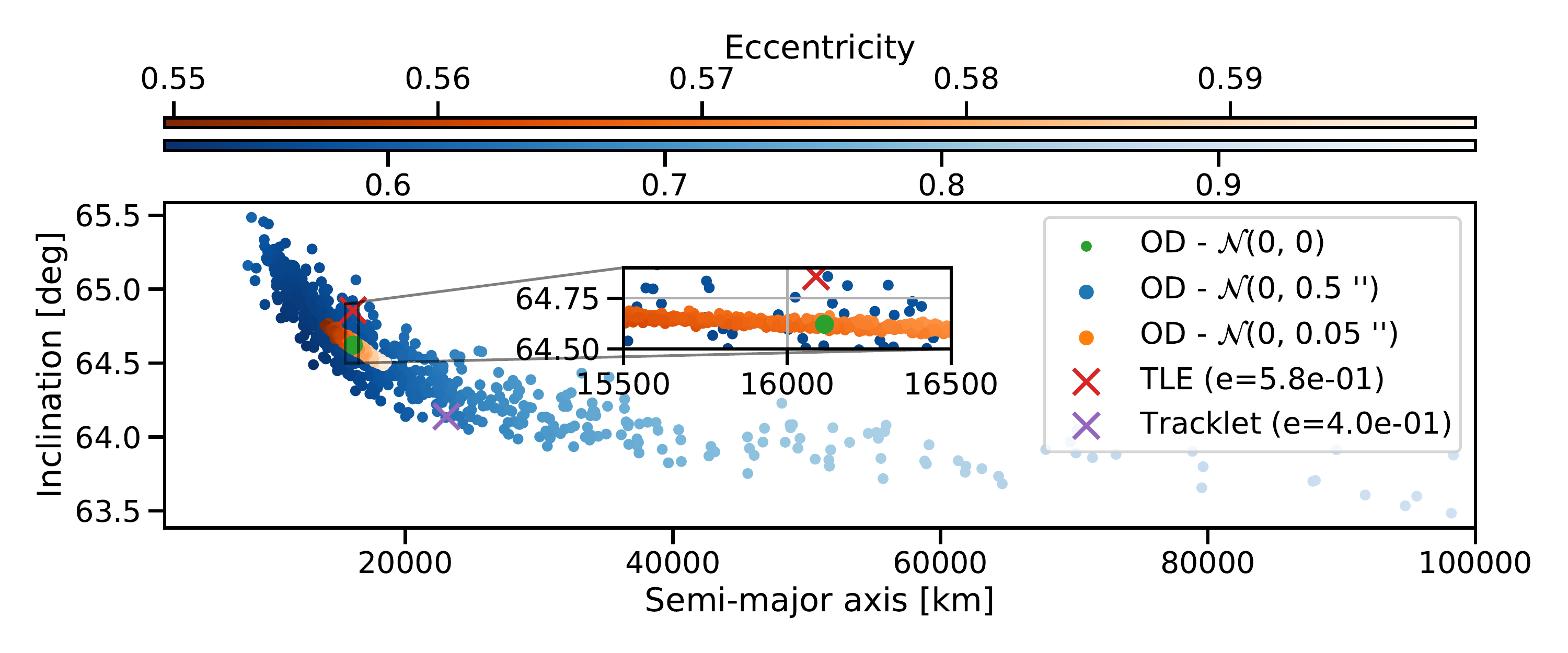}
    \includegraphics[width=1\linewidth]{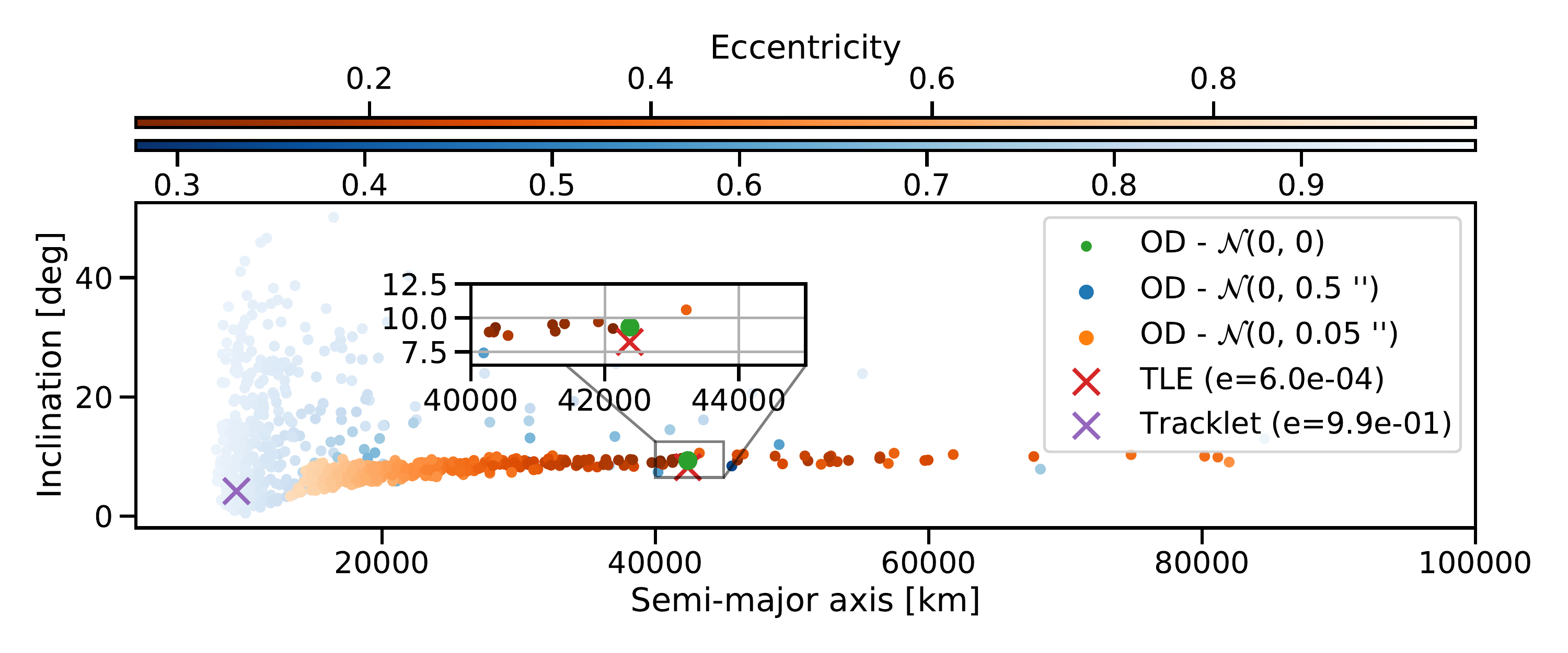}
    \caption{Impact of astrometric errors on tracklet orbital parameters estimation, for three types of tracklet. \textbf{Top:} data from ATLAS 5 CENTAUR DEB on 2021-09-09 03:32:32 UTC (93 events, arc length of 0.8 deg). \textbf{Middle:} data from SL-12 R/B (AUX MOTOR) on 2021-09-09 05:15:48 UTC (16 events, arc length of 1.0 deg). \textbf{Bottom:} data from GALAXY 9 on 2021-09-09 04:04:04 UTC (8 events, arc length of 0.1 deg). The orbital parameters from the TLE are shown with the red cross. The orbital parameters of the tracklet estimated using Orekit are shown with the purple cross. We also performed Orbit Determination (OD) using Orekit from perturbed TLE-generated measurements. 1,000 perturbations were drawn from a normal distribution centered on zero, with 0.5 arcsecond width (blue) or 0.05 arcsecond width (orange), and added to the propagated orbital elements of the satellite from the TLE. For reference, orbital parameters estimated directly from propagated orbital elements of the satellite without perturbations are shown with the green circle. See Sec. \ref{sec:tracklet-orbit} for more details.}
    \label{fig:orekit_orbits}
\end{figure}

Among the 6,450 tracklets, 2,609 (40\%) were not matched to objects from the NORAD catalogue. We tried to estimate their orbital parameters (semi-major axis $a$, inclination $i$, and eccentricity $e$) using Orekit\footnote{\url{https://orekit.org}. We used version 11.0.}, a low level and open source space dynamics library. 

There is no inherent timing information for individual alerts inside a tracklet -- they all are detected within the same 30-seconds exposure. We may (very roughly and in a generally biased manner) estimate these times by assuming that the first one is appeared exactly at the beginning of exposure, and the last one -- exactly at its end. Then we may simply linearly interpolate between the start and end of the exposure time using alerts spacing on the sky. Having this rough timing allows us to apply classical orbit determination methods to the tracklets.

\subsection{Orbit estimation using Orekit}\label{sec:orekit}

We extract the initial orbit parameters using the Laplace method \citep{bate1971fundamentals}. The method uses three RA/Dec measurements from the same observing site and timings to determine an initial orbit, assuming a Keplerian motion. In practice, we use the first, middle, and last alerts of each tracklet. Once the parameters of the initial orbit have been estimated, they are used to determine the full orbit. We use a batch least squares algorithm\footnote{\url{https://www.orekit.org/static/architecture/estimation.html}}. Due to the short arcs made by tracklets, we internally switch from Keplerian to Cartesian parameters to model the orbit for more stability.

Starting from a cross-identification of a tracklet with a satellite, we load the corresponding Two-Line Element (TLE) data and generate satellite positions for the duration of the exposure, at the estimated emission dates of the tracklet alerts. Figure \ref{fig:orekit_orbits} shows the Orekit solution based on this simulated data (green dot), and the orbital parameters from TLE data (red cross), for three different matches representing different types of tracklets: ATLAS 5 CENTAUR DEB on 2021-09-09 03:32:32 UTC (top panel, 93 events, arc length of 0.8 deg), SL-12 R/B (AUX MOTOR) on 2021-09-09 05:15:48 UTC (middle panel, 16 events, arc length of 1.0 deg), and GALAXY 9 on 2021-09-09 04:04:04 UTC (bottom panel, 8 events, arc length of 0.1 deg). Despite the small amount of information from tracklets in some cases (short arcs on the sky, few measurements per tracklet), we can get a correct estimate of the orbital parameters using Orekit assuming timings are correct.

In order to assess the validity for the time interpolation, we applied this procedure to all the tracklets matching with satellite data from the NORAD catalogue, and estimated the orbital parameters. For the vast majority of tracklets, the estimated parameters were very different from the reported TLE values. Estimated parameters often have very high eccentricity ($e>0.8$), with semi-major radius too small or too large by several factors.
The estimated inclinations are showing less discrepancy, but the differences are not negligible if one aims for precise measurements (up to a factor of two different). The Fig. \ref{fig:orekit_orbits} shows the parameters estimated from the tracklet data (purple cross), systematically far away from the parameters read directly from the TLE data (red cross), or the estimated parameters from an orbit determination using perfect satellite positions from the TLE data (green dot).

Then we tried to estimate more precisely the timings of the tracklets using the original ZTF images from the Data release 7, where we could see more than just valid alerts that pass the \fink\ quality cuts. We generated corresponding positions using TLE information for matching satellites. We found that most tracklet data span a time smaller than the exposure time. In most of the cases, the duration of the tracklet was estimated to be smaller than the exposure duration (one exposure is supposed to last for 30 seconds in total). In order to assess the residual in position, we rescaled our time interpolation and we shifted the data with respected to the first alert position (so that the first alert and the first TLE-generated positions are superimposed). By doing so the residuals in position for subsequent measurements were below 0.1 arcseconds (absolute values). Unfortunately, the rescaling in time did not improve the estimated orbital parameters, that were still largely discrepant with respect to the TLE data.

\subsection{Impact of astrometric accuracy}

To gain intuition on the impact of astrometric errors on orbital parameters estimation, we ran Orekit on TLE-generated data, with perturbations added to the input sky positions. Fig. \ref{fig:orekit_orbits} shows the recovered parameters for position perturbations drawn from a normal distribution centered on zero, and with widths of 0.5 arcsecond (blue dots) and 0.05 arcsecond (orange dots). The dots are color-coded by eccentricity values. For the figure, 1,000 perturbed measurements in each case have been generated, but in some cases Orekit was not able to recover the orbital parameters (hence there are less than 1,000 points on the figure in each case). We can clearly see degeneracy lines in each case, and the parameter space area spans grows for smaller tracklets. The parameters directly estimated from the tracklet data (purple cross) lie on the degeneracy, and correspond to RA/Dec perturbations around 0.05 arcsecond and above. We observed that for well sampled tracklets, or with big arcs (above 1 degree on the sky), the solution from tracklet data is less discrepant than for short or undersampled tracklets. Typically, in the case of short tracklets, the estimated eccentricity is systematically high ($e>0.8$), and the semi-major axis underestimated.

Timing errors happen to be critical to get a correct estimate of the orbital parameters. This translates to requirements on the sky positions of the order of $10^{-2}$ arcsecond, which seems difficult to achieve for wide field surveys -- even with the forthcoming Rubin Observatory. Alternatively, we could try to simultaneously estimate the timings and orbital parameters from position measurements. This approach, in the case of circular orbits ($e=0$), was developed in e.g. \cite{1973SvA....17..131Z}. While we could reproduce the results of the author using the example given in the paper, we could not apply successfully this method to our tracklet data (matching zero-eccentricity satellites), and more work would be necessary to understand the discrepancy. 

\section{Cataloguing and data availability}\label{sec:science-portal}

\begin{figure}
    \includegraphics[width=\linewidth]{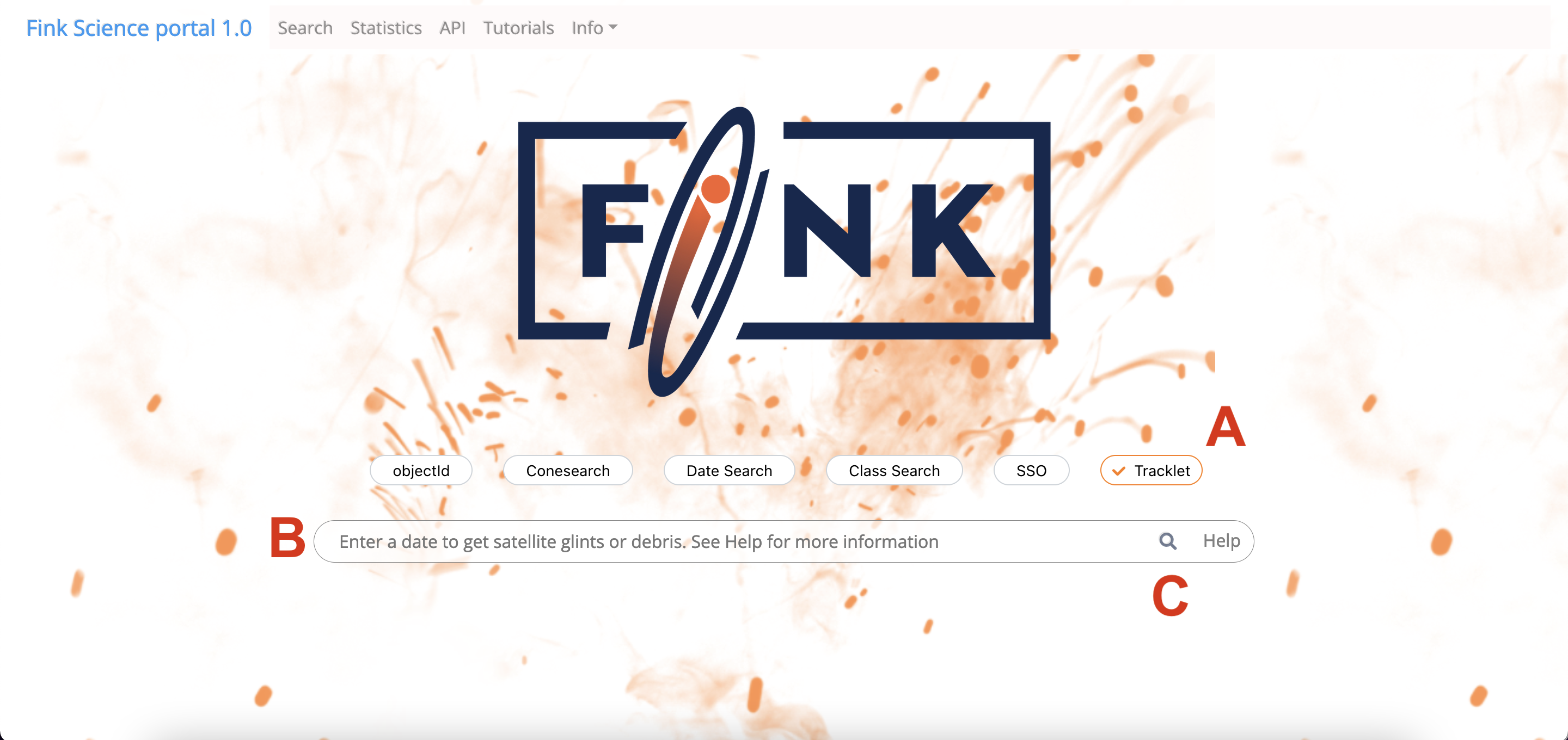}
    \caption{How to query for tracklet data on the \fink\ Science Portal. \textbf{A}: Select the \texttt{Tracklet} button. \textbf{B}: Enter a date at the format \texttt{YYYY-MM-DD hh:mm:ss}, or any subset. \textbf{C}: Hit search to display results. From the returned list of results, you can inspect entire tracklet data by clicking on one event, and hitting the tab \texttt{Tracklet}.}
    \label{fig:science-portal}
\end{figure}

Since November 2019, tracklets are being identified in the \fink\ alert stream and catalogued in its database, allowing to use this information in various kinds of searches. The tracklet extraction is done once per day, at the end of the observing night. The user can browse and extract the detected tracklet data from the \fink\ science portal  as shown in Fig. \ref{fig:science-portal}, or by directly using the REST API\footnote{See \url{https://fink-portal.org/api} for more details.}. For example, to query tracklet data from \fink\ using Python~3, the user needs to specify the date in UTC in the format \texttt{YYYY-MM-DD hh:mm:ss}\footnote{One can also specify bigger interval, e.g. \texttt{YYYY-MM-DD} to get all tracklets for one day, or \texttt{YYYY-MM-DD hh} to get all tracklets for one hour.}:

\begin{lstlisting}[language=Python, frame=single, caption={}, label=listing:tracklet]
import requests
import pandas as pd

# Get all tracklet data for the night 2021-09-09
r = requests.post(
  'https://fink-portal.org/api/v1/tracklet',
  json={
    'date': '2021-09-09',
    'output-format': 'json'
  }
)

# Format output in a Pandas DataFrame
pdf = pd.read_json(r.content)
\end{lstlisting}

Similarly, using the command-line tool \texttt{curl}:

\begin{lstlisting}[language=Bash, frame=single, caption={}, label=listing:curl]
curl -H "Content-Type: application/json" -X POST \
    -d '{"date":"2021-09-09", "output-format":"csv"}' \
    https://fink-portal.org/api/v1/tracklet -o trck_20210909.csv
\end{lstlisting}

This information is also immediately available for filtering out the non-astrophysical events while searching for alerts in \fink.

\bsp
\label{lastpage}
\end{document}